\documentclass[runningheads]{llncs}
\usepackage[T1]{fontenc}
\usepackage{csquotes}
\usepackage{graphicx}
\usepackage{cite}
\usepackage{caption}
\usepackage{subcaption}
\usepackage{setspace} 
\usepackage{multirow}
\usepackage{textcomp}
\usepackage{outlines}
\usepackage{booktabs}
\PassOptionsToPackage{hyphens}{url}
\usepackage[hidelinks]{hyperref}
\usepackage[demo,export]{adjustbox}
\usepackage[dvipsnames]{colortbl}
\usepackage{xcolor}
\usepackage{lscape}
\usepackage[margin=2.5cm]{geometry}
\def\BibTeX{{\rm B\kern-.05em{\sc i\kern-.025em b}\kern-.08em
    T\kern-.1667em\lower.7ex\hbox{E}\kern-.125emX}}
\usepackage{tcolorbox}

\newcommand{\Cc}{\textit{Command and Control }}
\newcommand{\vertical}[1]{\rotatebox[origin=r]{90}{#1}}

\definecolor{Gray}{gray}{0.9}
\definecolor{Dgray}{gray}{0.7}
\newcolumntype{g}{>{\columncolor{Gray}}c}
\newcolumntype{d}{>{\columncolor{Dgray}}c}
\newtcolorbox{takeaway}[2][]{%
fonttitle    = \bfseries,
title        = #1#2,
boxsep       = 0.5mm,
top          = 0.2mm,
bottom       = 0.2mm,
toptitle     = 0mm,
bottomtitle  = 0mm,
left         = 1mm,
right        = 1mm,
left skip    = 0mm,
right skip   = 0mm
}

\begin{document}
\title{Striking Back At Cobalt: Using Network Traffic Metadata To Detect Cobalt Strike Masquerading Command and Control Channels}
\titlerunning{Striking Back At Cobalt}

\author{Clément Parssegny\inst{1,2}\orcidID{0009-0004-2166-0881} \and Johan Mazel\inst{1}\orcidID{0009-0002-0222-6794} \and Olivier Levillain\inst{2}\orcidID{0000-0002-0558-5015} \and Pierre Chifflier\inst{1}}

\authorrunning{C. Parssegny et al.}
\institute{ANSSI, France\and SAMOVAR, Télécom SudParis, Institut Polytechnique de Paris, 91120 Palaiseau, France}

\maketitle 

\begin{abstract}
Off-the-shelf software for \Cc is often used by attackers and legitimate
pentesters looking for discretion.
Among other functionalities, these tools facilitate the customization of their
network traffic so it can mimic popular websites, thereby increasing
their secrecy.
Cobalt Strike is one of the most famous solutions in this category, used by
known advanced attacker groups such as "Mustang Panda" or "Nobelium".

In response to these threats, Security Operation Centers and other defense
actors struggle to detect \Cc traffic, which often use encryption protocols
such as TLS.
Network traffic metadata-based machine learning approaches have been proposed to detect
encrypted malware communications or fingerprint websites over Tor network.

This paper presents a machine learning-based method to detect
Cobalt Strike \Cc activity based only on widely used
network traffic metadata.
The proposed method is, to the best of our knowledge, 
the first of its kind that is able to adapt the model it uses 
to the observed traffic to optimize its performance.
This specificity permits our method to performs equally or better than the state of the art while using standard features.
Our method is thus easier to use in a production environment and more explainable.

\end{abstract}

\section{Introduction}
\label{section:introduction}

\Cc (C2) is used by attackers to control several compromised
hosts, forming networks called \textit{botnets}.
A usual architecture for these botnets is the client/server botnet.
In order to operate, this model of botnet uses communication channels where the
server sends commands to infected hosts that will execute them and then
transmit the results back.
These channels have evolved over the years.
First, they used IRC as their transport protocol.
Then, HTTP, HTTPS and DNS became commonly employed.
Recently, off-the-shelf commercial~\cite{bruteratel_website} or
open source~\cite{gsheetC2_github,manjusaka_github,sliver_github} software that were originally designed for Red team audits, have been used by malicious actors~\cite{bruteratel,gsheetC2,manjusaka,sliver}.

Cobalt Strike~\cite{CS_website} is the most popular example of these
frameworks~\cite{cobalt_strike_popularity}, used by both legitimate pentesters
and real threat actors~\cite{mustang_panda, nobelium}.
It can set up a full kill chain targeting Windows systems with several
attackers collaborating simultaneously~\cite{mitre_CS}.
This situation drives organizations to do everything they can to neutralize
identified Cobalt Strike-linked malicious
infrastructure~\cite{cobalt_strike_legal_action}.

In Cobalt Strike, \Cc communication channels can be customized to mimic
benign services, based on different configuration profiles.
Moreover, this off-the-shelf software supports TLS to encrypt
its communications.
Both masquerade and encryption strengthen Cobalt Strike's protection against detection.
However, the metadata related to the size, the direction or the timings of the
communication can still be leveraged for detection.

Botnet detection has been the goal of previous work through different ways.
Many used an approach based on machine
learning~\cite{Nivargi_botnet_2006,Livadas_botnet_2006,Kondo_c2_svm_2007,Dietrich_c2_dns_2011,
Warmer_c2_2011,Garcia_botnet_comparison_2014,Anderson_contextual_flow_2016,Pai_tls_signature_2020}.
However, as these Botnet \Cc channels rely on various protocols such as
IRC~\cite{Livadas_botnet_2006,Freiling_botnet_tracking_2005},
HTTP~\cite{Kondo_c2_svm_2007} or
DNS~\cite{Dietrich_c2_dns_2011, Anderson_contextual_flow_2016}, approaches used
for the detection vary.
While some research monitored a precise network's activity, measuring
anomalies as deviations from a computed 
profile~\cite{Cheung_grids_1998, Garcia_botnet_comparison_2014, Gu_botsniffer_2008},
others preferred to inspect packet contents to search for specific
characteristics of compromise~\cite{Nivargi_botnet_2006}.
As encryption use increased~\cite{Felt2017MeasuringHA}, network traffic
metadata began to be used to identify C2 traffic from benign
traffic~\cite{Pai_tls_signature_2020}.
However, up to our knowledge, malware targeted in existing works did not have 
a masquerade capability similar to Cobalt Strike.
A Cobalt Strike detection method~\cite{Ramos_ml_cobalt_strike_2022} has been proposed.
It is however limited regarding Cobalt Strike configurations and the training data 
is biased.
More recently, a deep learning approach has been presented~\cite{Yang_Petnet_2024}.
However, it does not take the masquerading capability of Cobalt Strike into account while using dubious feature construction steps.
In this paper, we focus on the capability for Cobalt Strike to obfuscate and
hide C2 traffic inside messages imitating benign services.
We address the existing work shortcomings and provide a detailed performance
evaluation based on configurations used in the wild.

Thus, our goal is to design a method to detect Cobalt Strike C2 traffic
despite its masquerade and encryption capabilities.
This method works with different configuration profiles and 
network protocols (namely HTTP, HTTPS and DNS).
Since configurations are easily modified and deployed in Cobalt Strike, it is
also central for our detection method to be easily extensible to seamlessly
include new configurations.
Our main contribution is a machine learning-based method to detect
Cobalt Strike C2 traffic using only network traffic’s metadata.
One advantage of this approach is that it is not based on Deep Packet Inspection
(DPI) at all but can still produce good performance, even with
unsophisticated and standard features.
Indeed, we observe mean $F_1$ scores between 0.78 and 1 with a confidence interval
at a 95\% threshold, equaling or slightly improving previous approaches.
These features also produce good performance on clear and encrypted Cobalt Strike C2 traffic from real-world attacks.

Our paper is structured as follows.
\autoref{section:related_work} describes existing work on C2 channel detection.
\autoref{section:cobalt_strike} portrays the working principle of Cobalt
Strike's \Cc feature.
\autoref{section:threat_model} presents the threat model we consider.
\autoref{section:method} depicts our method.
\autoref{section:results} details our results. 
Finally, \autoref{section:discussion} and \autoref{section:conclusion} discuss of 
our work and of our future work.

\section{Related Work}
\label{section:related_work}

C2 channel detection is inherently linked to botnet detection as it is
the communication method for such networks.
Decades ago, these botnets were based on IRC channels~\cite{Rajab_botnet_2006,Gu_bothunter_2007,Gu_botsniffer_2008} on which attackers would
send commands for victims to execute.
Then, HTTP~\cite{mitre_http_c2} and DNS~\cite{mitre_dns_c2} became widely
used protocols for C2.

Several papers searched for botnet by profiling the normal activity of the 
studied network before measuring the distance of a sample in comparison
with the benign one~\cite{Garcia_botnet_comparison_2014,Cheung_grids_1998,Gu_botsniffer_2008}.

Another method used is DPI~\cite{Bujlow_dpi_2015, Wang_payload_intrusion_detection_2004}.
The principle is to inspect payloads to detect botnet
messages~\cite{Nivargi_botnet_2006,
Gu_bothunter_2007} or to identify protocols~\cite{Gu_botsniffer_2008}
for correlation detection.
However, this method is only possible when the studied traffic is not
encrypted~\cite{Rajab_botnet_2006, Freiling_botnet_tracking_2005,
Livadas_botnet_2006, Nivargi_botnet_2006, Kondo_c2_svm_2007}
which is hardly the case now as the use of TLS increases continuously for 
both benign~\cite{Felt2017MeasuringHA} and C2 traffic.
Although methods like entropy computing~\cite{Dietrich_c2_dns_2011,
Zhang_encrypted_botnet_2013} were used to improve detection methods.

To accommodate the increasing use of encryption~\cite{Felt2017MeasuringHA},
papers that try to identify malicious traffic within encrypted communications
have been published.
They focus on machine learning techniques applied to TLS handshake
such as \texttt{ClientHello}, \texttt{Server\-Hello} or
\texttt{Certificate} messages~\cite{Pai_tls_signature_2020,
Anderson_contextual_flow_2016, Anderson_tls_fingerprint_2020,
Warmer_c2_2011}.
However, TLS~1.3, released in 2018,
encrypts most of the handshake messages~\cite{rfc_tls1.3}.

An overview of how machine learning has been applied for C2 detection
is presented in \autoref{tab:ml_state_of_art}.
It is noticeable that a majority of the research formalized this problem 
as a binary classification problem but then differ in the protocols studied.
Moreover, while other articles generally study one specific protocol in a known
configuration to evaluate and compare different machine learning methods,
our work concentrates on a unique method based on traffic metadata applied to a 
larger panel of configurations and protocols used in the wild.
We thus argue that the masquerading capability of Cobalt Strike C2 traffic
constitutes a unique challenge and we carefully design our proposed approach
to address this challenge.
Furthermore, few works detail the number of malicious configurations
they are studying, which has a detrimental influence on reproducibility and 
the generalization capability of their method.

\begin{table*}[tb]
	\centering
	\setlength\tabcolsep{5.8pt}
	\caption{Comparison of machine learning-based methods for C2 channels detection.
	Clas.: Classification type (B: Binary, M: Multiclass);
	Mw.: Malware;
	CS: Cobalt Strike;
	Feat. type: Feature Type (M: Metadata, D: Deep Packet Inspection);
	Reprod.: Reproducibility;
	Feat. imp. analysis: Feature Importance Analysis.
	Variability is the number of malicious classes used in the datasets e.g. the number of botnets or the number of malleable profiles.
	}
	\begin{tabular*}{\textwidth}{lglgcgcgcgcgcgcg}\toprule
  		\multicolumn{1}{l}{\textbf{Authors}} &
  		\multicolumn{1}{c}{\textbf{Ref.}} &
  		\multicolumn{1}{l}{\textbf{Yr.}} &
  		\multicolumn{1}{c}{\textbf{Clas.}} &
  		\multicolumn{4}{c}{\centering{\textbf{Protocols}}} &
  		\multicolumn{1}{c}{\multirow{2}{*}{\shortstack[c]{\textbf{Feat.}\\\textbf{type}}}} &
  		\multicolumn{2}{c}{\centering{\textbf{Target}}}&
  		\multicolumn{2}{c}{\centering{\textbf{Reprod.}}}&
  		\multicolumn{1}{c}{\multirow{2}{*}{\shortstack[c]{\textbf{Feat. imp.}\\\textbf{analysis}}}}\\
			
  		\cmidrule(lr){5-8}
  		\cmidrule(lr){10-11}
  		\cmidrule(lr){12-13}

  		 & & &
			&

  		\vertical{\textbf{IRC}} &
  		\vertical{\textbf{HTTP}} &
  		\vertical{\textbf{DNS}} &
  		\vertical{\textbf{TLS}} &
      &

  		\vertical{\textbf{Type}} &
  		\vertical{\textbf{Variability}}&
  		
  		\vertical{\textbf{Open code}}&
  		\vertical{\textbf{Open data}}\\

  		\midrule

		Nivargi et al.&\cite{Nivargi_botnet_2006} & '06 &
		B &
		$\surd$ & - & - & - &
		M/D & Mw. & ? &
		- & - & -\\

		Livadas et al. & \cite{Livadas_botnet_2006} & '06 &
		B/M &
		$\surd$ & - & - & - & 
		M & Mw. & 1 &
		- & - & -\\

		Kondo et al. & \cite{Kondo_c2_svm_2007} & '07 &
		B &
		$\surd$	& $\surd$ & - & - & 
		M & Mw. & 5 &
		- & - & - \\

		Dietrich et al. & \cite{Dietrich_c2_dns_2011} & '11 &
		B &
		- & - &	$\surd$ & - &
		M & Mw. & 11 &
		- & - & -\\

		Warmer & \cite{Warmer_c2_2011} & '11 &
		B &
		- & - & - &	$\surd$ & 
		M & Mw. & 4 &
		- & - & -\\

 		Garcia et al. & \cite{Garcia_botnet_comparison_2014} & '14 &
 		B &
 		$\surd$ &	$\surd$ & - & - &
 		M & Mw. &	10 &
 		- & $\surd$ & -\\

		Anderson et al. & \cite{Anderson_contextual_flow_2016} & '16 &
		B &
		- & $\surd$ &	$\surd$ & $\surd$ &
		M/D & Mw.& ? &
		-  & - & $\surd$\\

		Anderson et al. & \cite{Anderson_tls_fingerprint_2020} & '20 &
		M &
		- & - & - &	$\surd$ &
		M & Mw. & ? &
		$\surd$  & $\surd$  & $\surd$\\

 		Pai et al. & \cite{Pai_tls_signature_2020} & '20 &
 		B &
 		- & -	& - & $\surd$ &
		M & Mw. & ? &
		- & - & -\\

	  \cmidrule(lr){1-14}
  
		Van der Eijk et al. & \cite{VanDerEijk_Cobalt_strike_netflow_2020} & '20&
		B &
		- & $\surd$ & - & $\surd$ &
		M & CS & 1 &
		- & - & $\surd$\\
			
		Ramos et al. & \cite{Ramos_ml_cobalt_strike_2022} & '22 &
		B &
		- & $\surd$ & - & $\surd$ &
		M & CS & ? &
		- & - & - \\

		Ramos et al. & \cite{Ramos_ml_cobalt_strike_multiflow_2023} & '23 &
		B &
		- & - & - & $\surd$ &
		M & CS & 29 &
		- & - & - \\
		
		Yang et al. & \cite{Yang_Petnet_2024} & '24 &
		B &
		- & - & - & $\surd$ &
		M/D & CS & ? &
		$\surd$ & $\surd$ & - \\

 		\textbf{Our work}& & \textbf{'25}&
 		\textbf{B} &
 		\textbf{-}& $\surd$ & $\surd$ & $\surd$&
 		\textbf{M}& \textbf{CS} & \textbf{4} &
 		$\surd$ & $\surd$ & $\surd$\\
 		\bottomrule\\

	\end{tabular*}

	\label{tab:ml_state_of_art}
\end{table*}

It is also important to note that traffic metadata have also been used in active
methods to detect and fingerprint \Cc servers and infected hosts.
More specifically, TLS metadata present in the handshake messages were
studied by academics~\cite{Sosnowski_tls_fingerprint_2022, Sosnowski_dissectls_2023}
and industrial actors whose tools~\cite{ja3, jarm,ja4}
have been integrated to scanning platforms~\cite{censys} and are routinely used in CTI reports~\cite{jarm_cti}.
Finally, industrial research focused on detecting Cobalt Strike and
understanding its components to better fight against one of their client's
main threat~\cite{Talos_cobalt_strike_2020,BlackBerry_cobalt_strike_2021,ncc}.

Our method has common ground with the work of Anderson et
al.~\cite{Anderson_contextual_flow_2016} but we apply it to the Cobalt Strike
framework.
Van der Eijk et al.~\cite{VanDerEijk_Cobalt_strike_netflow_2020} designed
a threshold-based method to detect Cobalt Strike using
network traffic metadata.
However, there are limitations to their work.
First, they only detect one profile mimicking Amazon.
We address this issue by experimenting on four different profiles that mimic four distinct websites.
Then, they use empirically and manually chosen values as thresholds for their detection algorithm.
Finally, they only use the accuracy as performance metric.
As their data contains much more benign instances than malicious ones, 
their results are biased. 
We address this potential bias due to the dataset imbalance by using a
carefully designed training procedure and appropriate metrics.
More recently, Ramos et al.~\cite{Ramos_ml_cobalt_strike_multiflow_2023} used
multi-flow features to detect Cobalt Strike HTTPS sessions.
This approach has an objective close to ours, but their method focuses on a different scale of the communication.
Before this, Ramos et al.~\cite{Ramos_ml_cobalt_strike_2022} used supervised
machine learning to detect Cobalt Strike C2 traffic, but their work suffers
from several limitations.
First, they do not specify the number nor the details of the publicly
available and randomly generated malleable profiles used to train their
different models.
Hence, there is no assurance that their work is based on a realistic
deployment situation.
This also restricts the reproducibility of their results.
We address this issue by using profiles known to be deployed in the real 
world for our experiments and by documenting them.
Then, their dataset construction process suffers from some limitations.
They use both HTTP and HTTPS traffic for their malicious traffic while
only HTTPS traffic is used for the benign part.
This biases packet size-related network features.
We avoid this bias by using both HTTP and HTTPS for benign traffic.
We also take the DNS C2 use case into account while they do not.
Also, they transform categorical features into integers that encode the rank
in terms of decreasing appearance frequency in the dataset.
Thus the most frequent value is encoded as "1", the second most frequent value
as "2", etc.
This brings an order relation between initially unordered values and thus a bias for many methods such as tree-based Random Forest, which is put forward in the paper, or linear models such as Logistic Regression.
Thus, we argue that their model is too limited in terms of malleable profiles
and biased to be successful in a production environment.
Finally, they do not analyze the feature importance of their model,
as shown in~\autoref{tab:ml_state_of_art}.
This prevents from understanding why the model is efficient to detect 
Cobalt Strike C2 traffic.
We address this issue by computing the Mean Decrease in Impurity (MDI) of 
the different features used in our experiments.
More recently, Yang et al.~\cite{Yang_Petnet_2024} used a deep learning
approach to detect HTTPS traffic from Cobalt Strike.
Their approach suffers from several issues that we address in this work.
First, they are only taking HTTPS traffic into account.
By using one protocol, they omit several uses of Cobalt Strike C2 traffic.
We address this issue by proposing a method taking all the protocols usable to connect with a C2 server in Cobalt Strike.
Then, they do not consider the different malleable profiles they gathered in their datasets.
Moreover, by using only the default configuration in their C2 traffic generation lab,
they ignore the main difficulty regarding Cobalt Strike C2 traffic detection which is its masquerading capability and the associated configuration diversity~\cite{ncc}, as pictured in \autoref{subsection:data_collection}.
We address this issue by studying several configurations.
Furthermore, their approach based on deep learning may not be suitable in production environment as the detection is harder to explain.
Finally, some features, such as tokenized encrypted \textit{Application Data} payload exhibits dubious usefulness.
We address this issue by using common metadata features and explainable algorithms.

The detection of Cobalt Strike's C2 traffic can also fall within the more general
subject of detecting covert channels.
Our use case can be formalized as the detection of a Covert Storage Channel (CSC)
with a passive warden as presented in "The prisoner
problem"~\cite{simmons1984prisoners}.
This problem has been studied in several papers gathered in
surveys~\cite{Zander_covert_channels_2007, Mileva_covert_channel_tcp_ip2014}.
However, these detection techniques can not apply to covert channels with a
high degree of customization and an encryption of the application
layer which contains the payload, as in Cobalt Strike.
Machine learning approaches have also been
explored~\cite{Elsadig_ml_covert_channels_2022} but mainly for the DNS
protocol~\cite{Buczak_rf_in_pcap_2016} or timing based covert
channels~\cite{Chen_dns_covert_ml_2021}.
However, these detection techniques are making the hypothesis that
a single method and protocol are used by the covert channel and that
this information is known by the defender.
As Cobalt Strike is capable of using several protocols and methods to
masquerade its malicious traffic, we need a generic method able to adapt to 
different configurations.

\section{Cobalt Strike}
\label{section:cobalt_strike}

Cobalt Strike is an off-the-shelf penetration testing framework written in Java
and targeting Windows operating systems.
This section details the working principles of \Cc communications in
Cobalt Strike.
\autoref{subsection:Architecture} deals with the general architecture.
Then, \autoref{subsection:malleable_profiles} describes the different
possibilities brought by malleable profiles and how C2 protocols work.

\subsection{Architecture}
\label{subsection:Architecture}

\begin{figure}[b]
	\centering
	\includegraphics[width=0.9\textwidth]{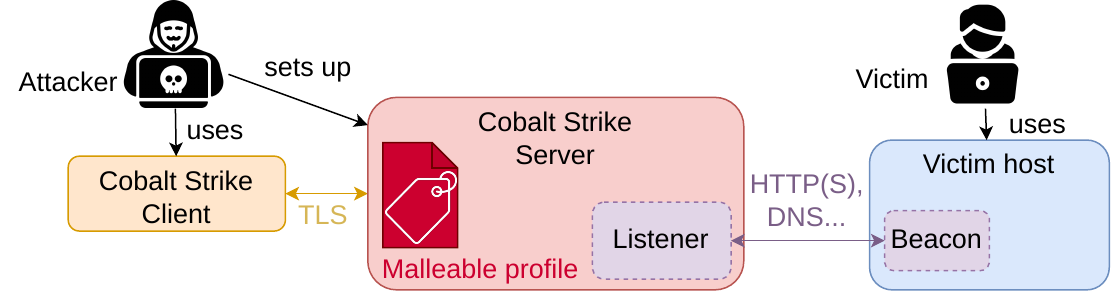}
	\caption{Cobalt Strike architecture.}
	\label{fig:C2_arch}
\end{figure}

\Cc in Cobalt Strike is based on three elements: a server controlled by
the attacker, a Beacon on each compromised system and a client used by the
attacker to control the server.
\autoref{fig:C2_arch} illustrates how these components interact with
each other.
The server is at the center of communications between attackers and
compromised systems.
It forwards commands from the attackers and perform on-demand initial exploitation executable generation.
In particular it creates Listeners and Beacons, the two agents of the C2
communication, based on the configured malleable profile.
Beacons are executed on a compromised system in order to execute C2 commands
selected by the attacker and sent by a Listener that is running on
the server side.
Finally, the client is a graphical interface used by the attacker to interact
with the server.

\subsection{Malleable Profiles}
\label{subsection:malleable_profiles}

\Cc configuration in Cobalt Strike is based on malleable profiles.
These profiles are files that define the behavior used during the
communication between the attacker server and the Beacons installed on
the target~\cite{CS_profiles_doc}.
A malleable profile contains all the parameters structuring
the generated traffic, in particular the metadata used to impersonate
benign traffic such as the values of the HTTP headers parameters, the encoding
method used to obfuscate the payload or the "sleeptime" between two C2
communications.
The malleable profile may also define a TLS certificate
as Cobalt Strike can use an existing or self-signed certificates to create
HTTPS C2 channels.
As multiple servers can use the same profile
(see \autoref{subsection:data_collection}), detecting a profile may allow one
to identify several servers that use similar malleable profiles.

Communication behaviors can be defined for different protocols within
a single profile as Cobalt Strike supports HTTP, HTTPS, DNS, SMB and TCP.
In the following, we focus only on HTTP(S) and DNS protocols because SMB and
TCP are only used in a peer-to-peer connection between two Beacons and not
between a Beacon and a server.
This Beacon-only communication is designed to limit
the number of hosts calling out to the server during
lateral movements~\cite{cobaltstrike_p2p}.

The C2 communication can be described in four steps.
First, at configurable time intervals (with or without jitter),
the Beacon contacts the Listener on the server with identification metadata 
to check if there are commands to execute.
This is called a "check-in".
Then, the Listener responds with the attacker's commands or,
if there is none, with the default answer defined in the profile.

\begin{figure*}[tb]
	\centering
	\begin{subfigure}[b]{0.4\textwidth}
		\includegraphics[width=\textwidth]{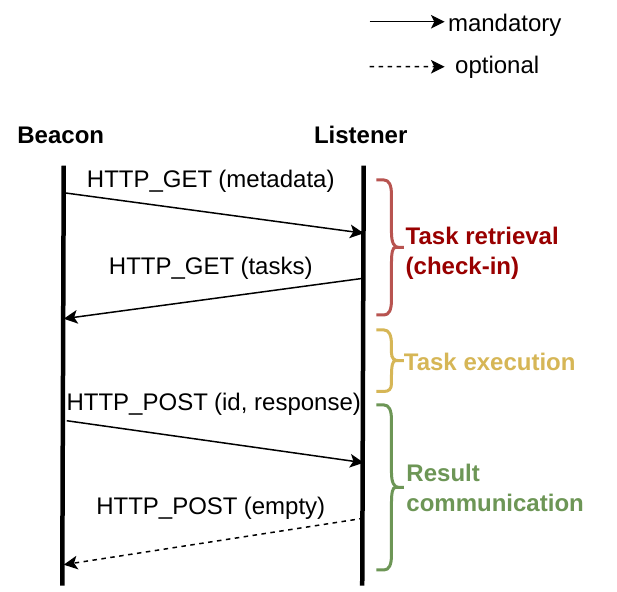}
		\caption{HTTP }
		\label{fig:http}
	\end{subfigure}
	\hspace{0.1\textwidth}
	\begin{subfigure}[b]{0.38\textwidth}
		\includegraphics[width=\textwidth]{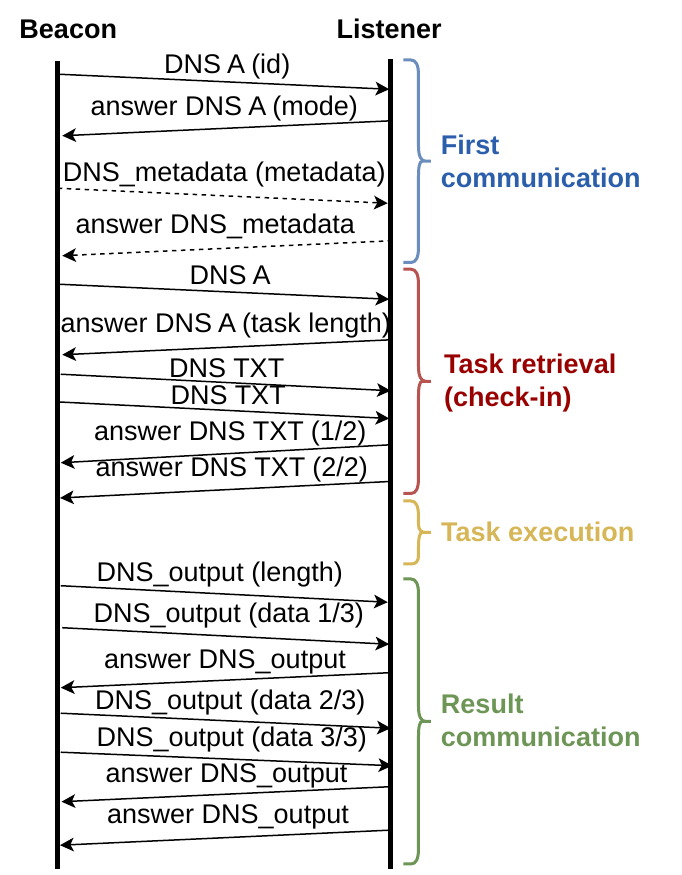}
		\caption{DNS in TXT mode.}
		\label{fig:dns}
	\end{subfigure}
        \caption{Examples of C2 communications based on HTTP and DNS.}
  \label{fig:C2_protocols}
\end{figure*}

Once the output of the commands is known, the Beacon sends it to the server
with a different request.
This request also contains authentication metadata, so the server can sort
the requests when several Beacons are used.
Finally, the Listener can send a final answer which is optional.
For HTTPS, a TLS handshake is executed to fetch the orders and post
the outputs.
The version of TLS used depends on the OpenJDK version used by Cobalt Strike.
The DNS communication is slightly more complex than the HTTP(S) one
but follows the same logic.
An illustration of these protocols is presented in \autoref{fig:C2_protocols}.

\section{Threat model}
\label{section:threat_model}

We consider the following threat model.
The attacker controls an external C2 server and at least one compromised
machine inside the supervised network, running a Cobalt Strike Beacon.
Thus, these two hosts establish C2 channels using either HTTP, HTTPS or DNS.
We, as defenders, can passively monitor incoming and outgoing network traffic,
but we cannot analyze the application layers. 
To put it in a covert channel detection context, 
as described by Simmons~\cite{simmons1984prisoners}, we are a passive warden
trying to detect a Covert Storage Channel between the Cobalt Strike server and
the infected host which are the two prisoners.

Our goal is then to distinguish C2 communications from benign traffic.

\section{Method}
\label{section:method}

We detail the proposed method in the following parts.
\autoref{subsection:principle} outlines its general principle.
\autoref{subsection:data_collection} presents our data generation processes
while \autoref{subsection:dataset_construction} depicts the feature extraction
of the metadata for the datasets' construction.
\autoref{subsection:ml} describes our machine learning method.

\subsection{Principle}
\label{subsection:principle}

Our objective is to detect Cobalt Strike C2 communications,
using only network traffic metadata.
This method should work with different malleable profiles and should be 
extensible to easily take into account new malleable profiles.

The use of malleable profiles makes it complex to detect
Cobalt Strike C2 channels using patterns in headers or payload.
The encryption of the application layers introduced by recent changes in security
protocols, such as TLS~1.3, DNS over HTTPS or encrypted \texttt{Client Hello}
(formerly limited to an encrypted \texttt{Server Name Indication}) also adds to the detection obstacles regarding packet
content inspection.
Supervised machine learning models based on metadata are thus an answer
to masquerade and encryption.

We use a supervised machine learning model for each "group" of malleable profiles we consider, each profile in a "group" mimicking the same benign traffic.
This facilitates the extension of the proposed method to new profiles.
Our method aims to select the most relevant model regarding the
context to obtain the best performance.
Thus, with our method, a single model may help detect slightly different 
malleable profiles, for example derived from the public repository~\cite{cobalt_strike_github}.

\begin{figure*}[tb]
	\centering
 	\includegraphics[width=\textwidth]{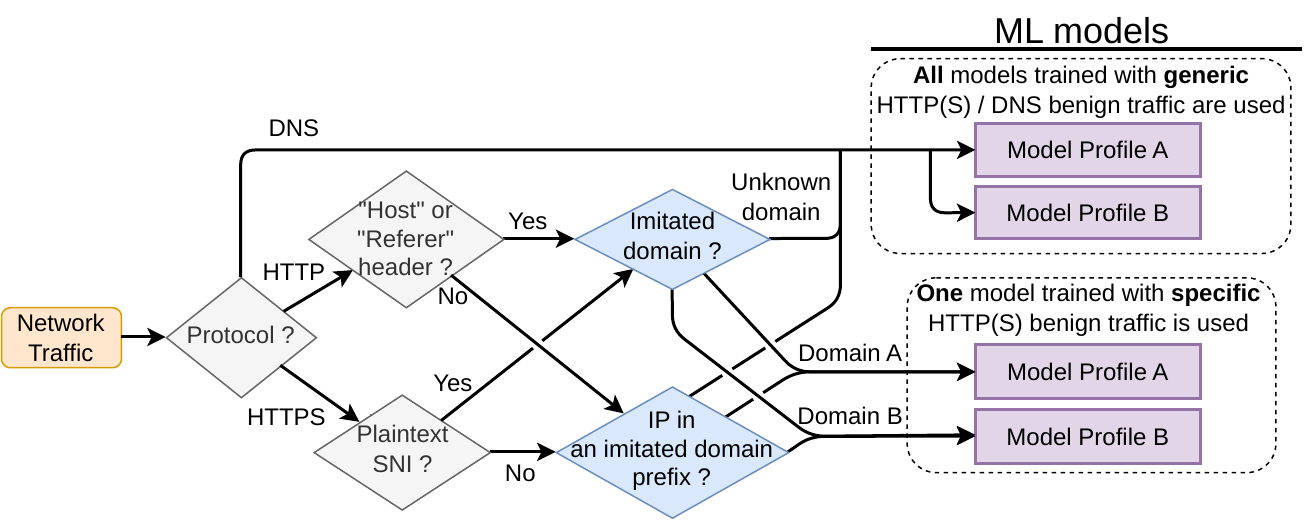}
	\caption{Proposed method to detect Cobalt Strike C2 traffic.
    We use machine learning models trained on traffic which uses the same 
    protocol (HTTP, HTTPS or DNS) as the observed traffic.
	Each model is trained using benign traffic as indicated on the figure and malicious traffic generated using the related profile.}
	\label{fig:detection}
\end{figure*}

The proposed method is depicted in \autoref{fig:detection}.
First, we make the hypothesis that we know what protocol is used by the traffic
to be classified (e.g. based on the UDP/TCP ports).
Because the purpose of Cobalt Strike is to be stealthy and overcome
firewalls, we suppose attackers use the same ports as benign traffic.
Then, we check if a domain name is available in the traffic metadata.
This information may be present in different locations such as the "Host"
header for HTTP or the \textit{Server Name Indication} for TLS.
If we cannot extract a domain name,
we check if the remote IP address belongs to a subnet part of a
domain present in a known malleable profile (such as \texttt{3.0.0.0/15} which
belongs to Amazon).
If the observed traffic is linked to a known domain, we apply a machine
learning model trained with malicious and benign traffic specific
to this domain.
Otherwise, we use several models, each trained with domain-specific malicious 
traffic and generic benign traffic.

\subsection{Data collection}
\label{subsection:data_collection}

We set up two platforms to generate network traffic.
The first one is based on a virtualized architecture to run \Cc scenarios and
capture the malicious traffic generated by the attack.
The second one uses an automation framework to mimic user
activity and create benign traffic as close as possible to real traffic.
We also use external benign traffic datasets to minimize bias.
The details of the data used in our experiments are listed in
\autoref{tab:nb_instances}.

\begin{table*}[tb]
	\centering
	\setlength\tabcolsep{23.4pt}
	\caption{Number of instances (TCP or UDP flow) used in our experiments.
  }
	\begin{tabular*}{0.9\textwidth}{lccr}\toprule
  
	  \multicolumn{1}{l}{\textbf{Traffic Type /}}&
	  \textbf{Protocol}&
	  \textbf{(B)enign /}&
	  \multicolumn{1}{r}{\textbf{Instance}}\\

	  \multicolumn{1}{l}{\textbf{Malleable Profile}}& & \textbf{(M)alicious}&\multicolumn{1}{r}{\textbf{Count}}\\\midrule

	  \multirow{3}{*}{\shortstack[l]{Generic traffic\\(UPC+UPNA+CTU)}} & HTTP & B & 73 873\\
	  & HTTPS & B & 74 019\\
	  & DNS & B & 279 455\\
	  \cmidrule(lr){1-4}

	  \multirow{2}{*}{Default profile} & HTTPS & M & 247\\
	  & DNS & M & 7 618\\
	  \cmidrule(lr){1-4}

	  \multirow{4}[4]{*}{Amazon} & HTTP & M & 172\\
	  \cmidrule(lr){2-4}
	  & \multirow{2}{*}{HTTPS} & M & 160\\
	  & & B & 615\\
	  \cmidrule(lr){2-4}
	  & DNS & M & 23 507\\
	  \cmidrule(lr){1-4}
  
	  \multirow{2}{*}{Smashburger} & \multirow{2}{*}{HTTPS} & M & 129\\
	  &  & B & 230\\
	  \cmidrule(lr){1-4}

	  \multirow{2}{*}{jQuery} & \multirow{2}{*}{HTTP} & M & 2 225\\
	  & & B & 99\\
	  \cmidrule(lr){1-4}

	   MTA Jan. 31st~\cite{mta_2023-01-31} & HTTPS & M & 158\\
	   MTA May 23th~\cite{mta_2023-05-23} & HTTPS & M & 121\\
	   MTA Jul. 12th~\cite{mta_2023-07-12} & HTTPS & M & 803\\
	   MTA Oct. 3rd~\cite{mta_2023-10-03} & HTTPS & M & 203\\
	   MTA Nov. 6th~\cite{mta_2023-11-06} & HTTP & M & 2 208\\
		\bottomrule\\
	\end{tabular*}

	\label{tab:nb_instances}
\end{table*}

\subsubsection{Malicious traffic}
\label{subsubsection:malicious_traffic}

Because Cobalt Strike is a tool designed for Windows attacks, we use
a Windows Server 2022 virtual machine (VM) as the victim.
The remaining part of the Vagrant-based architecture is composed of two
Debian~11 VMs.
One is the Cobalt Strike server, and the other is a \texttt{bind9} server used in DNS-based C2 channels experiments.

Our detection approach targets network traffic observed between hosts (i.e. not
on the host themselves).
TCP offloading mechanisms are thus deactivated with the \texttt{ethtool} package
to respect the default Ethernet \textit{Maximum Transmission Unit} (MTU) value
of 1 500 bytes.
C2 commands, listed in \autoref{tab:cmd}, are chosen to be diverse
regarding the amount of generated traffic and provided functionality.

\begin{table}[tb]
	\centering
	\setlength\tabcolsep{13.8pt}
	\caption{C2 commands used for malicious traffic generation.}
	\begin{tabular*}{0.7\textwidth}{ll}\toprule
  		\textbf{Command} & \textbf{Action}\\\midrule
		\texttt{bhashdump} & Dump local passwords hashes\\
	    \texttt{blogonpassword} & Dump passwords with Mimikatz\\
	    \texttt{bls} & List current directory\\
	    \texttt{bmimikatz} & Use a Mimikatz command\\
	    \texttt{bnet} & Run a network module command\\
	    \texttt{bportscan} & Scan the network using
		\textit{Beacon}'s port scanner\\
	    \texttt{bps} & List processes\\
	   	\texttt{bpwd} & Print current directory path\\
	    \texttt{brun} & Run a console command\\
		\texttt{bpowershell} & Run a console command (brun variant)\\
	    \texttt{bscreenshot} & Take and send a screenshot\\\bottomrule
	\end{tabular*}
	\label{tab:cmd}
\end{table}

We also gather malicious Cobalt Strike C2 traces from \href{https://www.malware-traffic-analysis.net/}{Malware Traffic Analysis} (MTA)~\cite{mta} to test our method and features on real-world attacks.

\subsubsection{Benign traffic}

The benign traffic generation process is based on the Python bindings to the
Selenium framework.
This tool is used to simulate web browsing-related user inputs.
Therefore, we write a script adapted to each service spoofed by Cobalt Strike
that we study.
For example, the script for Amazon search and select items based on
predefined search terms as the corresponding studied Cobalt Strike profile
uses "item search" like requests.
Meanwhile, we capture the generated network traffic.
We also use generic traffic from external datasets from 
the Universitat Politècnica de Catalunya (UPC)~\cite{Bujlow_dpi_2015, upc_dataset},
the Universidad Pública de Navarra (UPNA)~\cite{Labayen_ml_user_activities_2020, upn_dataset}
and the Stratosphere IPS project of the Czech Technical University (CTU)~\cite{stratodatasets} 
to minimize the bias of the benign traffic dataset.
To use this traffic with our malicious data and minimize bias again, only flows
that respect our condition on MTU are kept as stated
in \autoref{subsubsection:malicious_traffic}.
It is then split by protocol we want to study: HTTP, HTTPS or DNS.

\subsubsection{Choice of targeted malleable profiles}

Cobalt Strike provides its own malleable profile set~\cite{cobalt_strike_github}
with an official repository containing 32 malleable profiles based on popular
services and known attackers.
Because this dataset does not provide any information about its use in attacks, our choice of malleable profiles is based on the
Beacon dataset published by NCC~\cite{ncc}.
It contains more than 120~000 Cobalt Strike Beacons fetched in the wild between
2018 and 2022.

We extract every profile in the dataset based on the IP/port pair and the month
of its collection to avoid duplicates.
After determining the domain mimicked for each of them by looking at the headers parameters, we apply a fuzzy
hashing algorithm, TLSH, to get a ranking of the most used profiles~\cite{tlsh}.
We also use DBSCAN to cluster profiles with small differences, even if the mimicked domain value is different.
DBSCAN parameters are $\epsilon=30$ and $minPts = 2$.
Here, the $\epsilon$ parameter is the maximal distance between two TLSH digests as defined in~\cite{tlsh} within a cluster.
Based on this information, we select three domains:
\texttt{amazon.com}, \texttt{code.jquery.com} which are the two most mimicked
domains and \texttt{smashburger.com} which has a significant low number of
unique profiles for a great number of instances.
The 10 most mimicked domains are listed on \autoref{tab:profils}.
Finally, for each of these three chosen websites, we select the most
common profile that we adapt so the C2 server answers with realistic answers.
For example the jQuery profile returns the legitimate code after receiving a
command result.
The hash of the profiles used are listed in \autoref{tab:profile_hash}.

\begin{table}[tb]
	\centering
	\setlength\tabcolsep{4.8pt}
	\renewcommand{\arraystretch}{0.85}
	\caption{Top 10 most common domains mimicked in the Cobalt Strike's Beacons
	dataset published by NCC group~\cite{ncc}. Each Beacon is counted once per month per
	(IP, port) couple. Nb: Number.}
 	\begin{tabular*}{0.9\textwidth}{lrrr}\toprule
		\multicolumn{1}{p{.15\textwidth}}{
			\raggedright \textbf{Spoofed domain}
		}&
		\multicolumn{1}{p{.24\textwidth}}{
			\centering \textbf{\textbf{Nb of TLSH-based grouped profiles}}
		}&
		\multicolumn{1}{p{.15\textwidth}}{
			\centering \textbf{Nb of instances}
		}&
		\multicolumn{1}{p{.15\textwidth}}{
			\raggedleft \textbf{Nb of \\unique profiles}
		}\\\midrule
		
		code.jquery.com & 30 & 4 949 & 601\\
		www.amazon.com & 36 & 3 662 & 243\\
		download.windowsupdate.com & 42 & 993 & 145\\
		locations.smashburger.com & 2 & 827 & 3\\
		www.google.com & 40 & 722 & 195\\
		www.bing.com & 29 & 548 & 92\\
		ocsp.verisign.com & 24 & 520 & 54\\
		www.microsoft.com & 33 & 360 & 67\\
		onedrive.live.com & 15 & 316 & 63\\
		audio-sv5-t1-3.pandora.com & 7 & 281 & 38\\\bottomrule
	\end{tabular*}
	\label{tab:profils}
\end{table}

\begin{table}[tb]
	\centering
	\setlength\tabcolsep{31.8pt}
	\caption{Hashes of the different malleable profiles used.}
	\begin{tabular*}{\textwidth}{ll}\toprule
  		\textbf{Malleable Profile} & \textbf{SHA-1 hash}\\\midrule

		Default & \texttt{cb8632399e2c07b7b69e2403f3d543ac870176a4}\\
		Amazon & \texttt{e3ee5e42845ef28a19fd6dd39b418acab279582d}\\
	    jQuery (with realistic answers) & \texttt{9b6551fb41c96b47f3e4153a0603458166fe44c6}\\
	    Smashburger & \texttt{1561b087573c535fc953336b509df88de82b75ba}\\	    
			\bottomrule
	\end{tabular*}
  	\label{tab:profile_hash}
\end{table}

As many web services now enforce the use of HTTPS, the jQuery profile is chosen for the HTTP experiments.
Indeed, this service is still accepting HTTP requests while others enforce HTTPS connections.
For the HTTPS experiments, we use self-signed certificates generated by Cobalt Strike with mimicked
parameters such as the Common Name for more stealthiness.
Additionally, we use the default profile for testing HTTPS and DNS detection.
This profile is found in several hundreds instances in the NCC dataset but with much variations, making it difficult to establish the correct count.

\subsection{Dataset construction}
\label{subsection:dataset_construction}

Once the network traffic is captured, we extract the metadata
to build the dataset.
This is done by using the Zeek analysis framework~\cite{zeek}.
Our Zeek scripts fetch the metadata for each flow, i.e. each TCP or UDP
connection, from an input network capture.
We also make sure to remove flows without any exchanged data, e.g. from a port scanning, as it would bring bias in the learning process.

Netflow is a protocol developed by Cisco to collect information and statistics on an IP network traffic based on \textit{flows}.
A \textit{flow} is described using different values that differ between the versions of the protocol.
We choose Netflow v5~\cite{cisco:netflow_v1578} and 
Netflow v9~\cite{cisco:netflowv9} features for our experiments.
We limit our work on this metadata as they are standard and easy
to collect in a production environment.
We also extend these 2 sets with features computed on Netflow information to improve the classification process.
Finally, we compute two ratios.
One is the ratio of the total size of packets received to the total size of
packets sent while the other is the same logic applied to
the number of packets.
Ramos et al.~\cite{Ramos_ml_cobalt_strike_2022} uses three other features from a standard Zeek log~\cite{zeek_conn_log}.
The first is the ordered history of the TCP flags received in a flow.
The second is the transport protocol used (TCP or UDP) while \textit{service} is an inference on the application protocol based on the destination port e.g. 53 for DNS.
These feature groups are presented in \autoref{tab:fg}.

\begin{table*}[tb]
	\centering
	\setlength\tabcolsep{10.35pt}
    	\renewcommand{\arraystretch}{1.2}
	\caption{Used network traffic features.
	ext.: extended;
	$\mathcal{B}$: bi-direction (Beacon to Listener and Listener to Beacon
	jointly) ;
	$\mathcal{U}$: uni-directional (Beacon to Listener and Listener to Beacon
	separately) ;
	$\mathcal{A}$: all directions (bi-direction and two uni-directions) ;
	R: $\frac{received}{sent}$ ratios ;
	S: sent packets only (Beacon to Listener);
	L3/4: Layer 3/4.
	Byte counts are based on the layer payload.
	}
	\begin{tabular*}{\textwidth}{lcccccccc}\toprule
        \textbf{\shortstack[c]{Feature\\group name}}&
        \multicolumn{8}{c}{\centering \textbf{\shortstack[c]{Flow\\ information}}}\\
    
        \cmidrule(lr){2-9}

		& \vertical{\textbf{Nb packets}} &
		\vertical{\textbf{Packet size}} &
		\vertical{\textbf{\centering\shortstack[c]{OSI layer\\ used for size}}}&
		\vertical{\textbf{Duration}} &
 		\vertical{\textbf{TCP flags}} &
  	\vertical{\textbf{TCP history}} &
  	\vertical{\textbf{Protocol}} &
  	\vertical{\textbf{Service}}

	   \\\midrule
      
		Netflow v5 & $\mathcal{B}$ & total ($\mathcal{B}$) & L3 & $\surd$ & $\surd$ & - & - & -\\
		Netflow v5 ext.& $\mathcal{B}$ & total \& mean ($\mathcal{B}$)& L3 &$\surd$ & $\surd$& - & - & -\\
		\multirow{2}{*}{Netflow v9} &  \multirow{2}{*}{$\mathcal{U}$} &  total ($\mathcal{U}$);&  \multirow{2}{*}{L3} & \multirow{2}{*}{$\surd$} &  \multirow{2}{*}{$\surd$}&  \multirow{2}{*}{-} &  \multirow{2}{*}{-}&  \multirow{2}{*}{-}\\
		 \addlinespace[-1ex]
			& & minimum \& maximum (S) & & & & &\\
			
	    \multirow{2}{*}{Netflow v9 ext.}& \multirow{2}{*}{$\mathcal{A}$; R} & total
	    \& mean ($\mathcal{A}$); & \multirow{2}{*}{L3} &\multirow{2}{*}{$\surd$}& \multirow{2}{*}{$\surd$}& \multirow{2}{*}{-} & \multirow{2}{*}{-} & \multirow{2}{*}{-}\\
\addlinespace[-1ex]
	    & & minimum \& maximum (S); R & & & & &\\

	    Ramos et al.~\cite{Ramos_ml_cobalt_strike_2022} & $\mathcal{B}$ & total ($\mathcal{B}$)& L4& - & -& $\surd$ &$\surd$&$\surd$\\\bottomrule\\
	\end{tabular*}
	\label{tab:fg}
\end{table*}

\subsection{Machine Learning: pre-processing, algorithm and metrics}
\label{subsection:ml}

After constructing the datasets, we apply supervised machine learning
classification techniques.
First, we scale the features using standardization.
Then, we run a Random Forest algorithm, which is known to give good results and
understandable explanation compared to other methods such as
deep learning-based ones.
We use a stratified k-fold cross-validation with a common value
of $k = 10$ for the learning process to limit the bias from the imbalance
between the proportion of benign and malicious traffic in the dataset.
Finally, we use a grid search for hyperparameter tuning and optimize 
our performance.
This grid search tunes the number of trees (10, 100 or 500),
the quality of a split criterion (gini or entropy),
the depth achievable by a tree (2, 5, 10, 15 or 20) and
the minimum of sample for a split to occur (2, 5, 10 or 50).

Because we are in a security use case where alerts are processed by
human beings, it is important to limit the number of false positives.
Thus, we choose the $F_1$ score as a metric.
We also use precision and recall metrics to better analyze
the performances of the models. 
To measure the importance of each feature in the decision of the model,
we use the Mean Decrease in Impurity (MDI).

\begin{table*}[tb]
	\centering
	\setlength\tabcolsep{17.75pt}
	\caption{Mimicked activities and network traffic used in experiments.
	The labels of the experiments associated are listed in the last column.
	JS : JavaScript,	
	Smashb.: Smashburger.}
 \begin{tabular*}{\textwidth}{lccr}\toprule
	
	\textbf{\shortstack[l]{Mimicked\\activity}} &
	\textbf{\shortstack[c]{Targeted malleable\\profile traffic (protocol)}} &
	\textbf{\shortstack[c]{Benign traffic\\(protocol)}} &
	\textbf{\shortstack[r]{Experiment\\label}}
	\\\midrule
	
	\multirow{7}{*}{\shortstack[l]{Website\\browsing}} & \multirow{2}{*}{Amazon
	(HTTP)} & Amazon
	(HTTPS) & Az/Az (HTTP)\\
	& & Generic (HTTP) & Az/Gen (HTTP)\\
	\cmidrule(lr){2-4}
	& \multirow{2}{*}{Amazon (HTTPS)} & Amazon (HTTPS) & Az/Az (HTTPS)\\
	& & Generic (HTTPS) & Az/Gen (HTTPS)\\
	\cmidrule(lr){2-4}
	& \multirow{2}{*}{Smashb. (HTTPS)} & Smashb. (HTTPS) & Sb/Sb (HTTPS)\\
	& & Generic (HTTPS) & Sb/Gen (HTTPS)\\
	\cmidrule(lr){2-4}
	& Default (HTTPS) & Generic (HTTPS) & D/Gen (HTTPS)\\
	\midrule	
	
	\multirow{2}{*}{\shortstack[l]{JS library\\download}} &
	\multirow{2}{*}{jQuery~3.6 (HTTP)} &	jQuery (HTTP) & jQ/jQ (HTTP)\\
	& & Generic (HTTP) & jQ/Gen (HTTP)\\
	\midrule
	
	\multirow{2}{*}{DNS} & Default (DNS) & \multirow{2}{*}{Generic (DNS)} &
	D/Gen (DNS)\\
	& Amazon (DNS) & & Az/Gen (DNS) \\
	\bottomrule\\
	\end{tabular*}    

	\label{tab:exp}
\end{table*}

\section{Results}
\label{section:results}
The goal of these experiments is to evaluate the performance of
the models used in our method which are depicted in purple boxes 
in~\autoref{fig:detection}.
We compare them to the Random Forest approach of Ramos et al.~\cite{Ramos_ml_cobalt_strike_2022} which is put forward in their article as it has a good $F_1$ score and the smallest false positive rate.
Yang et al.~\cite{Yang_Petnet_2024} is not considered as deep learning is not easily explainable.

\begin{figure*}[tb]
	\centering
	\setlength\tabcolsep{3.3pt}
	\begin{tabular*}{\textwidth}{lccccccc}
		&\multicolumn{2}{c}{\textbf{DNS}} & \multicolumn{2}{c}{\textbf{HTTP}} & \multicolumn{3}{c}{\textbf{HTTPS}}\\

		\cmidrule(lr){2-3}
		\cmidrule(lr){4-5}
		\cmidrule(lr){6-8}

		&\textbf{Default} & \textbf{Amazon} & \textbf{Amazon} & \textbf{jQuery} & \textbf{Default} & \textbf{Amazon} & \textbf{Smashb.}\\		

	  \rotatebox{90}{\parbox{3cm}{\centering\textbf{Generic}}} &
		\includegraphics[height=3cm]{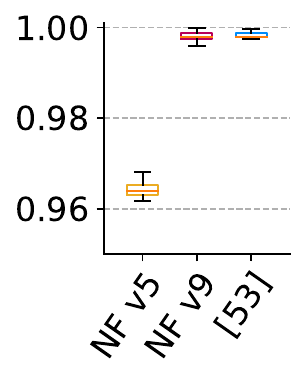} &
  	\includegraphics[height=2.95cm]{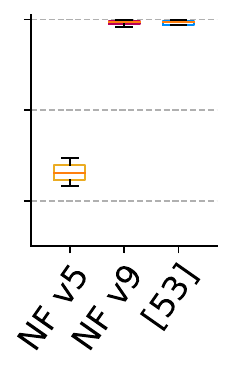} &
		\includegraphics[height=3cm]{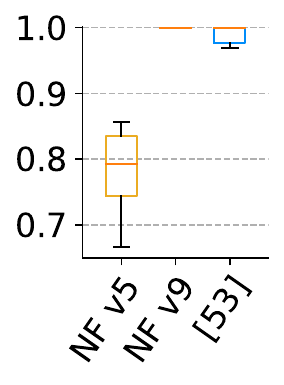} &
		\includegraphics[height=2.9cm]{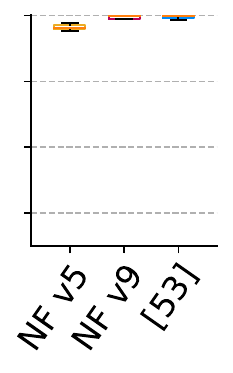}&
		\includegraphics[height=3cm]{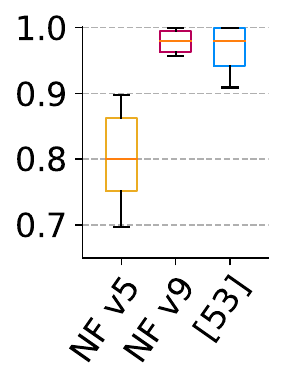} &
		\includegraphics[height=2.9cm]{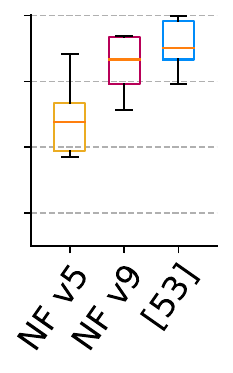} &
		\includegraphics[height=2.9cm]{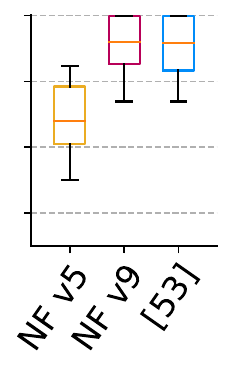}\\
				
  	& \textbf{(a)} & \textbf{(b)} & \textbf{(c)} & \textbf{(d)} & \textbf{(e)} & \textbf{(f)} & \textbf{(g)} \\

		\rotatebox{90}{\parbox{3cm}{\centering\textbf{Mimicked}}} &
			&
			&
		\includegraphics[height=3cm]{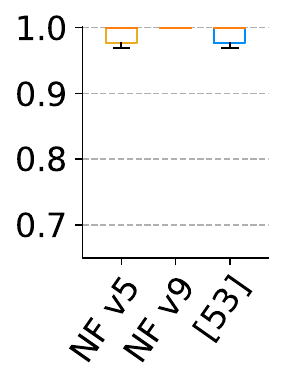} &
		\includegraphics[height=2.9cm]{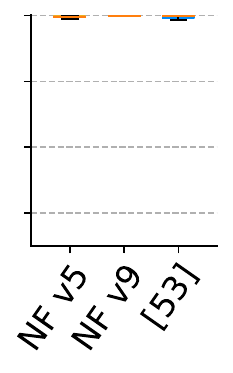}&
			&
		\includegraphics[height=3cm]{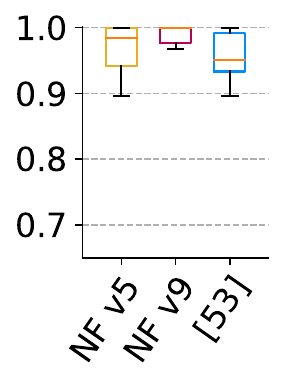} &
		\includegraphics[height=2.9cm]{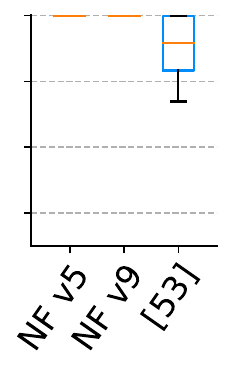}\\
				
		& & & \textbf{(h)} & \textbf{(i)} & & \textbf{(j)} & \textbf{(k)} \\

	\end{tabular*}
	\caption{$F_1$ score of the different models.
	Ramos et al.~\cite{Ramos_ml_cobalt_strike_2022} uses generic DNS or HTTP and HTTPS combined
	benign traffic.
	Smashb.:~Smashburger,
	NF:~NetFlow
  }
	\label{fig:boxplots}
\end{figure*}

The different sets of features and malleable profiles used are
depicted in \autoref{tab:fg} and \autoref{tab:exp}.
Netflow uses the layer 3 payload size, designated in the following as "size".
To visualize the results, we generate box plots of the F1 scores of
the different models in~\autoref{fig:boxplots}.
Each box plot is based on the 10 values from the 10 cross-validation folds.

Our preliminary experiments show that the duration feature has a strong
impact on the model's performance.
This may be explained by the fact that our malicious server and victim are
located in a virtualized environment where there is less RTT and jitter
whereas the benign traffic is generated using servers in the wild.
We also observe a strong impact of CWR and ECE TCP flags, explained by
the fact that benign traffic datasets are generated using Linux operating
systems which disable these flags by default, whereas the malicious traffic is necessarily produced by Windows hosts where these flags are enabled by default.
To limit this bias, we perform experiments without the duration or these
two flags.

We present our results in the following parts.
First, \autoref{subsection:detection_DNS_traffic} deals with the case of
DNS C2 traffic.
Then, \autoref{subsection:detection_unknown_domain} compares the performances
of the two methods when no known profile can be linked to the observed traffic.
\autoref{subsection:detection_mimicked_domain} analyzes the results
when a known domain can be linked to the traffic observed.
Finally, in \autoref{subsection:detection_in_the_wild}, we apply our method
and compare the performances of Netflow features to those
Ramos et al.~\cite{Ramos_ml_cobalt_strike_2022} propose using publicly shared malicious traces~\cite{mta_2023-01-31, mta_2023-05-23, mta_2023-07-12, mta_2023-10-03, mta_2023-11-06}.

As we observe that the extended feature sets only slightly improve the performance, we focus on the basic sets which are easier to collect in
a production environment.
Thus we compare three methods: our method with Netflow v5 features, our method with Netflow v9 features (both without the features discarded in the previous paragraph), and Ramos et al.~\cite{Ramos_ml_cobalt_strike_2022} method.

\begin{figure*}[tb]
	\centering
	\setlength\tabcolsep{1.2pt}
	\begin{tabular*}{\textwidth}{lccc}
		& 	\includegraphics[width=0.16\textwidth, valign=c]{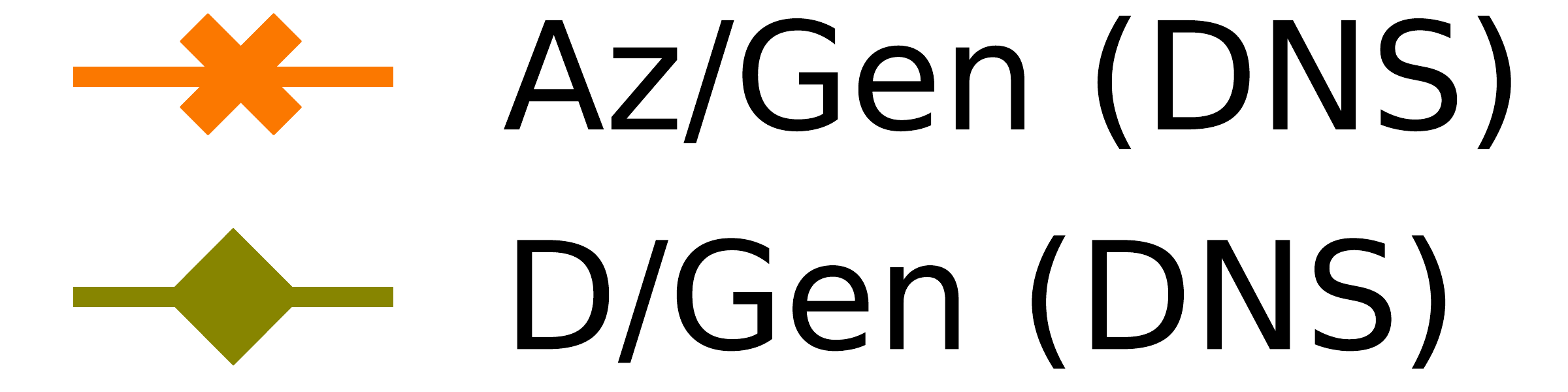}
		&  	\includegraphics[width=0.35\textwidth, valign=c]{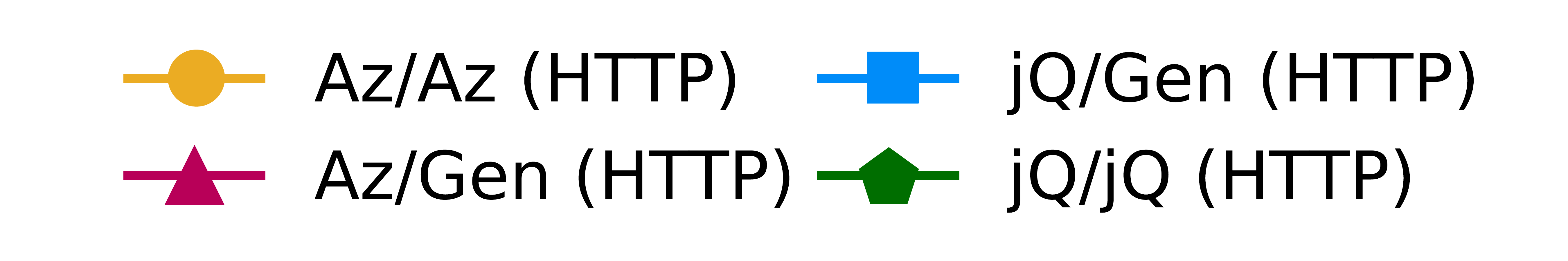}
		&  	\includegraphics[width=0.3\textwidth, valign=c]{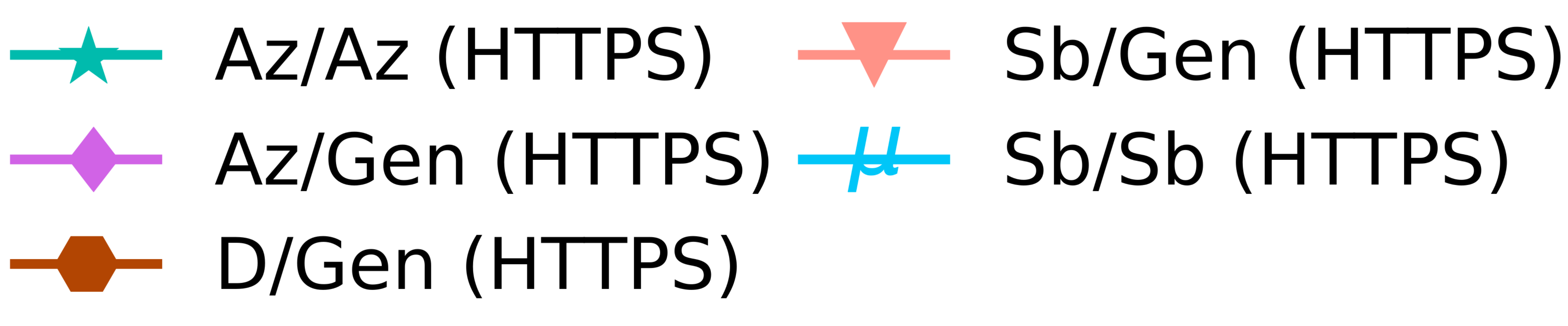}\\

		& \textbf{DNS} & \textbf{HTTP} & \textbf{HTTPS}\\

		\rotatebox{90}{\parbox{3cm}{\centering\textbf{Ramos et al.}}} &
		\includegraphics[height=3cm]{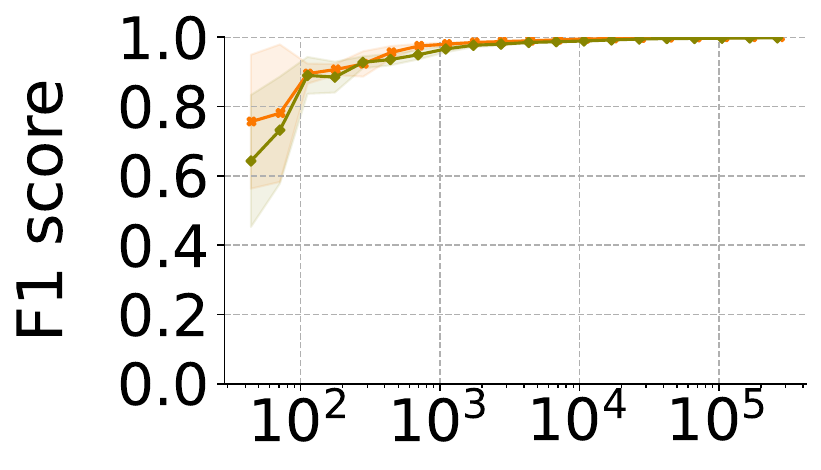} &
		\includegraphics[height=2.8cm]{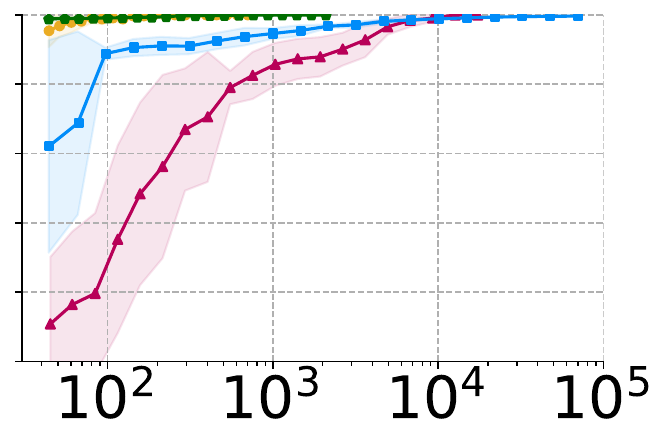}&
		\includegraphics[height=2.8cm]{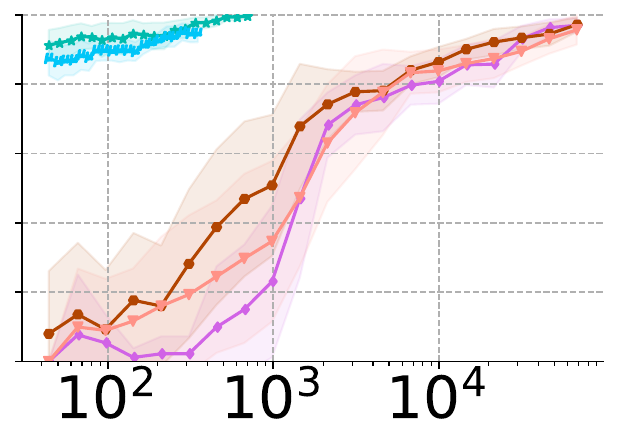}\\

		& \textbf{(a)} & \textbf{(b)} & \textbf{(c)}\\

	  \rotatebox{90}{\parbox{3cm}{\centering\textbf{NF v5}}} &
		\includegraphics[height=3cm]{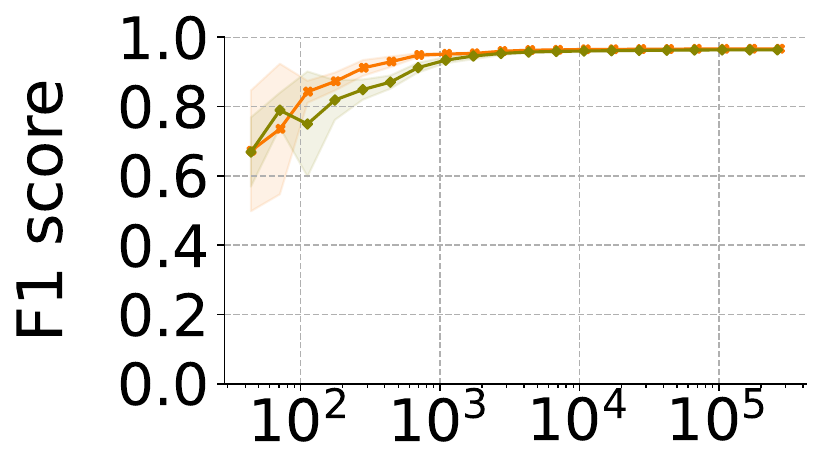} &
		\includegraphics[height=2.8cm]{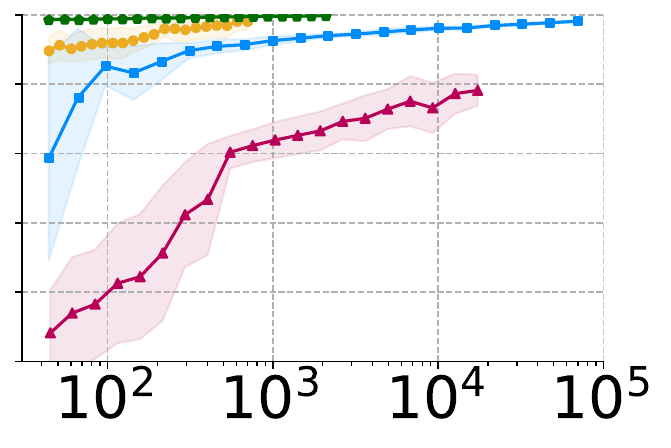}&
		\includegraphics[height=2.8cm]{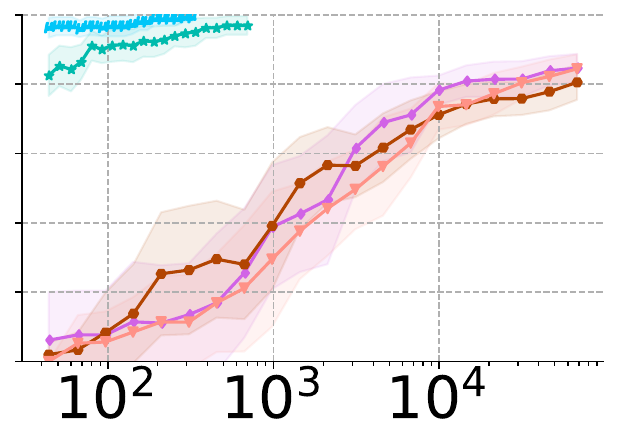}\\

		& \textbf{(d)} & \textbf{(e)} & \textbf{(f)}\\

		\rotatebox{90}{\parbox{3cm}{\centering\textbf{NF v9}}} &
		\includegraphics[height=3cm]{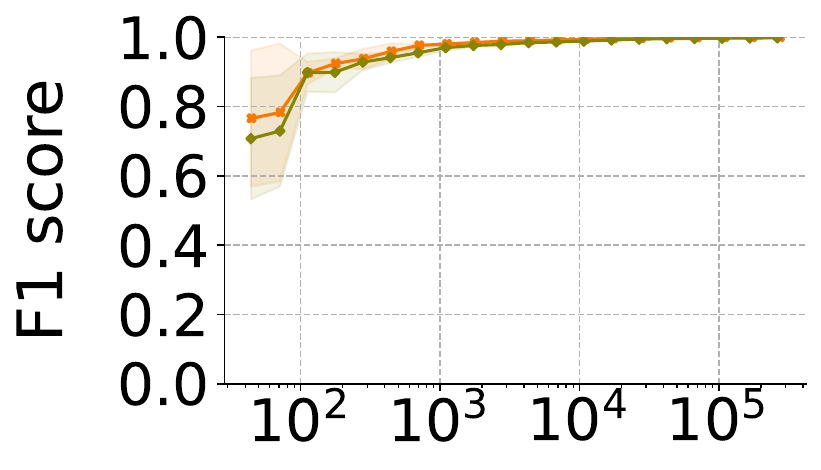} &
		\includegraphics[height=2.8cm]{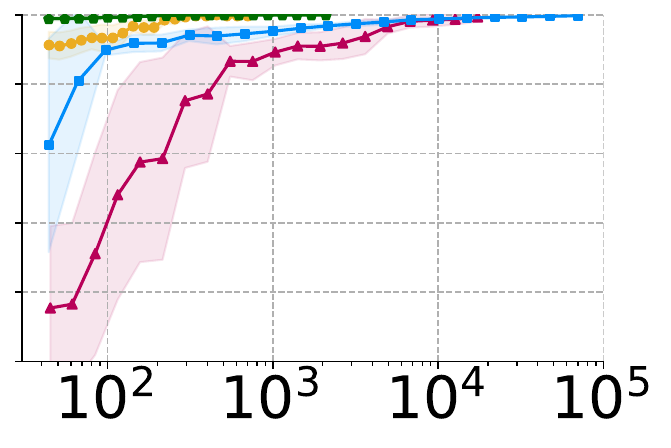}&
		\includegraphics[height=2.8cm]{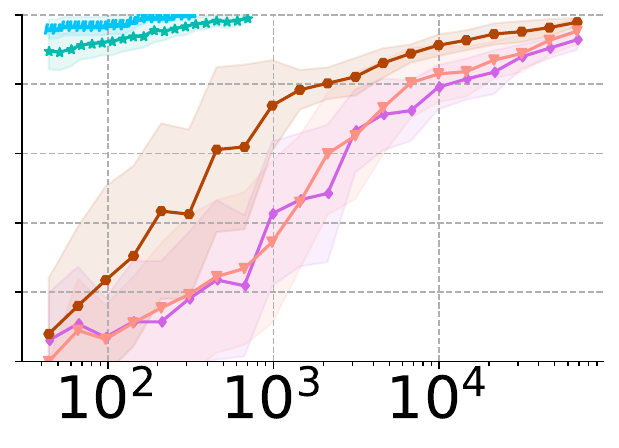}\\

		& \textbf{(g)} & \textbf{(h)} & \textbf{(i)}\\

		& \textbf{Nb training flows} & \textbf{Nb training flows} & \textbf{Nb training flows}\\

	\end{tabular*}
	\caption{Learning curves using $F_1$ score with a confidence interval
	at a 95\% threshold for the different feature groups.
	NF:~NetFlow,
	Az:~Amazon,
	D:~default,
	Gen:~generic,
	jQ:~jQuery,
	Sb:~smashburger.}
	\label{fig:lc_noduration}
\end{figure*}

To visualize the results, we also generate the learning curves with
a confidence interval at a 95\% threshold, as presented in~\autoref{fig:lc_noduration}.
Then, we group the experiments by protocol used in the C2 traffic.
The learning curves are composed of twenty points corresponding to training
sizes equally spaced on a logarithmic scale between fifty samples and the size
of the overall training dataset.
The stratified k-fold cross-validation discussed before guarantees
the separation between the training and testing flows for each point of the curve.
Since some of the cases studied generate more flows than others, e.g. web
browsing vs jQuery downloading, the sample sizes used for the learning curves differ.

\begin{figure*}[tb]
	\centering
	\setlength\tabcolsep{1.55pt}
	\begin{tabular*}{\textwidth}{lccc}
		& 	\includegraphics[width=0.17\textwidth, valign=c]{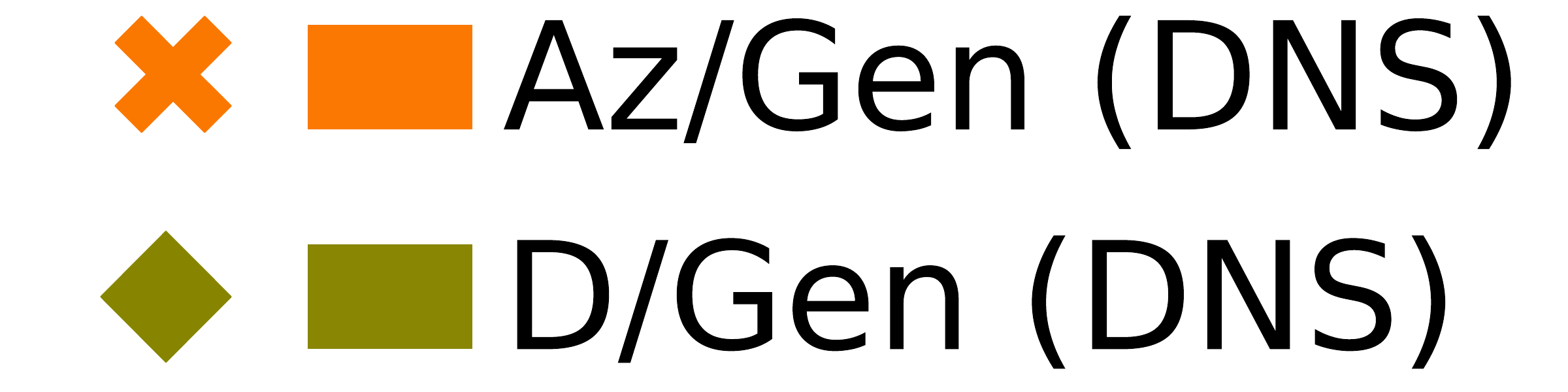}
		&  	\includegraphics[width=0.34\textwidth, valign=c]{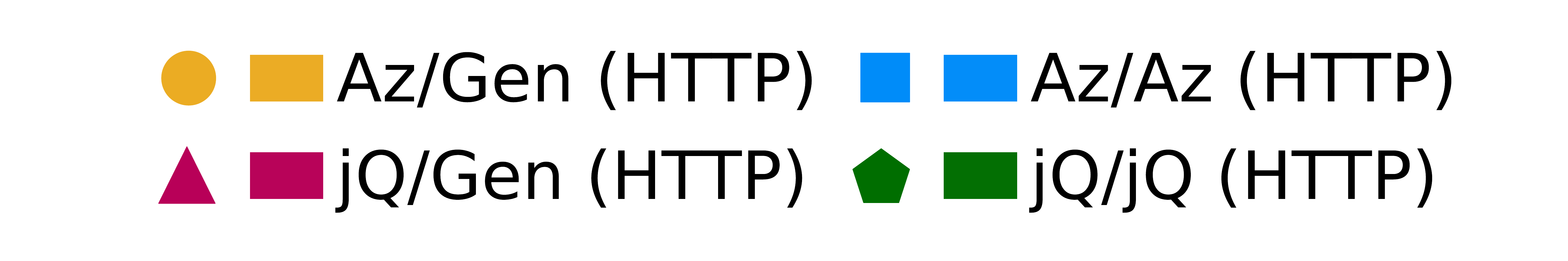}
		&  	\includegraphics[width=0.34\textwidth, valign=c]{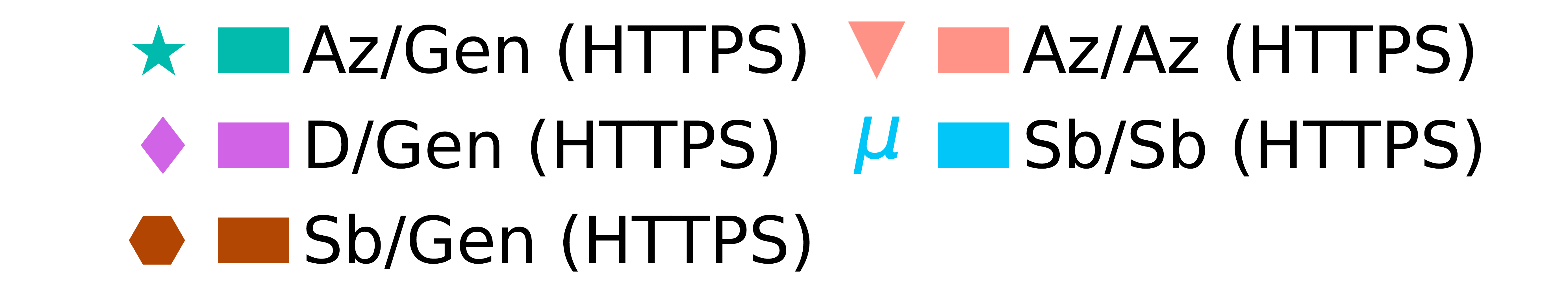}\\
			&	\small\textbf{DNS}
			& \small\textbf{HTTP}
			& \small\textbf{HTTPS}\\

			\rotatebox{90}{\parbox{4.6cm}{\centering{\textbf{Generic}}}}&
			\includegraphics[height=4.5cm]{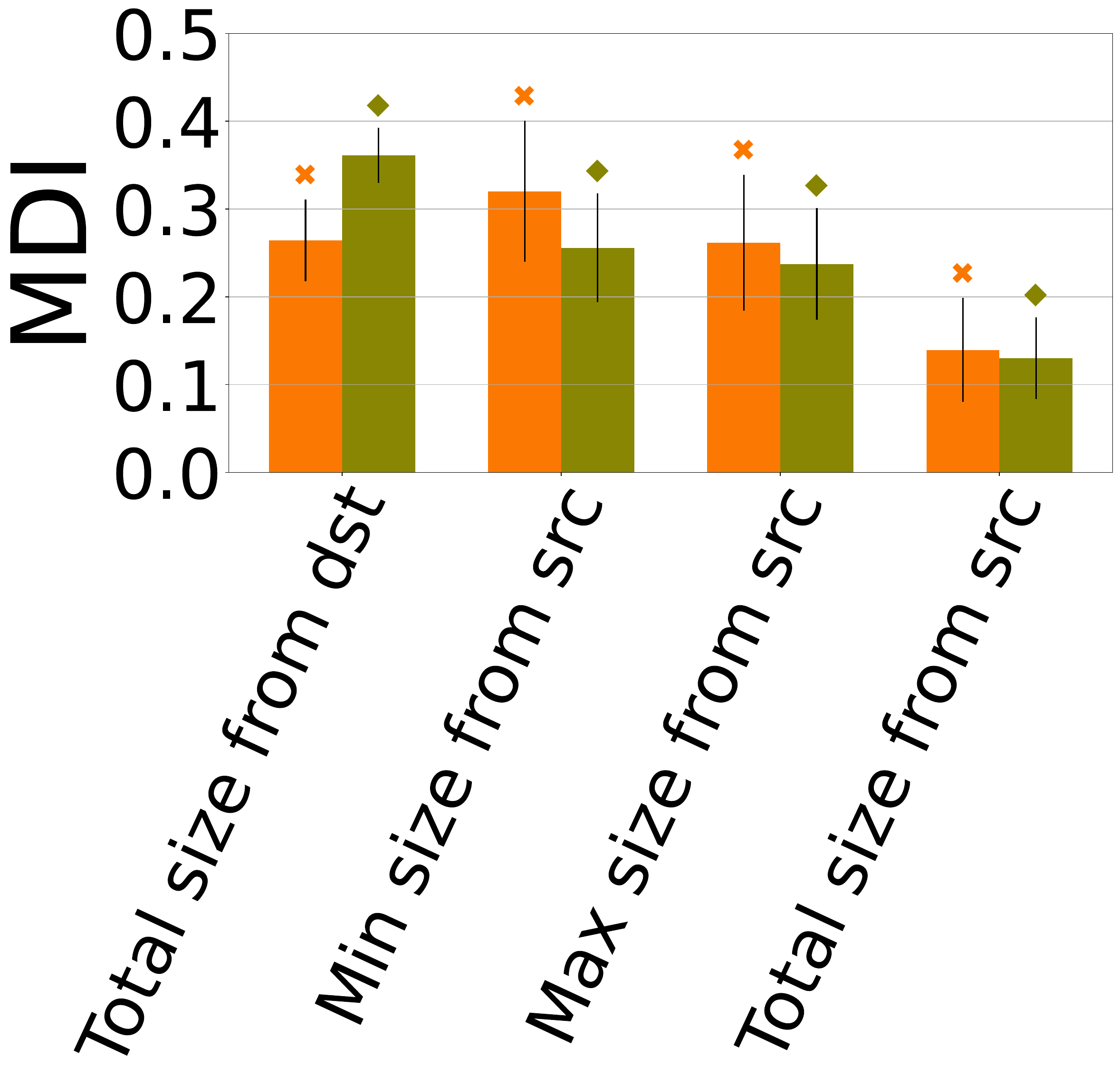}&
			\includegraphics[height=4.5cm]{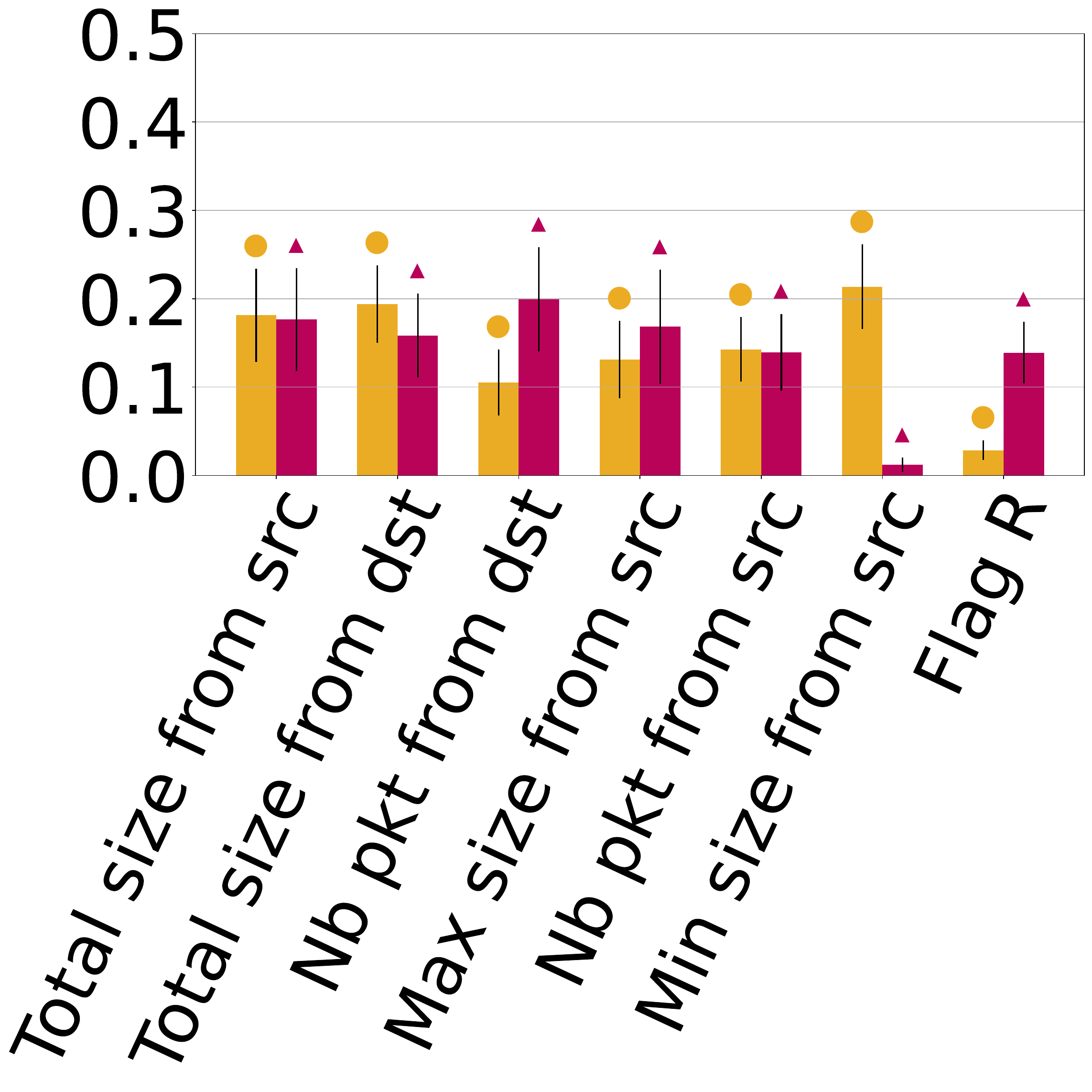}&
			\includegraphics[height=4.5cm]{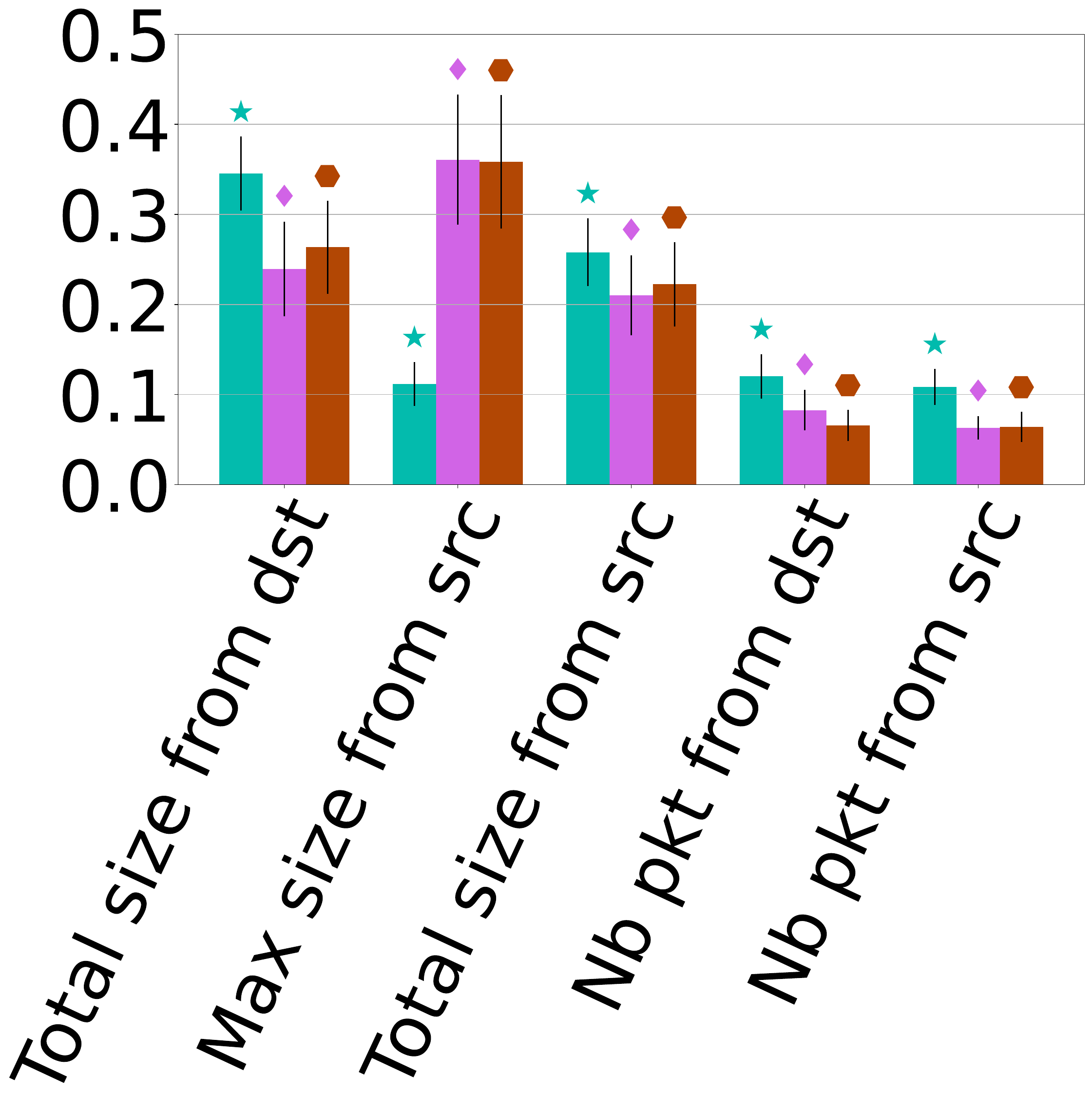}\\

		& \textbf{(a)} & \textbf{(b)} & \textbf{(c)}\\
			
			\rotatebox{90}{\parbox{4.5cm}{\centering{\textbf{Mimicked}}}}&
			&
			\includegraphics[height=4.5cm]{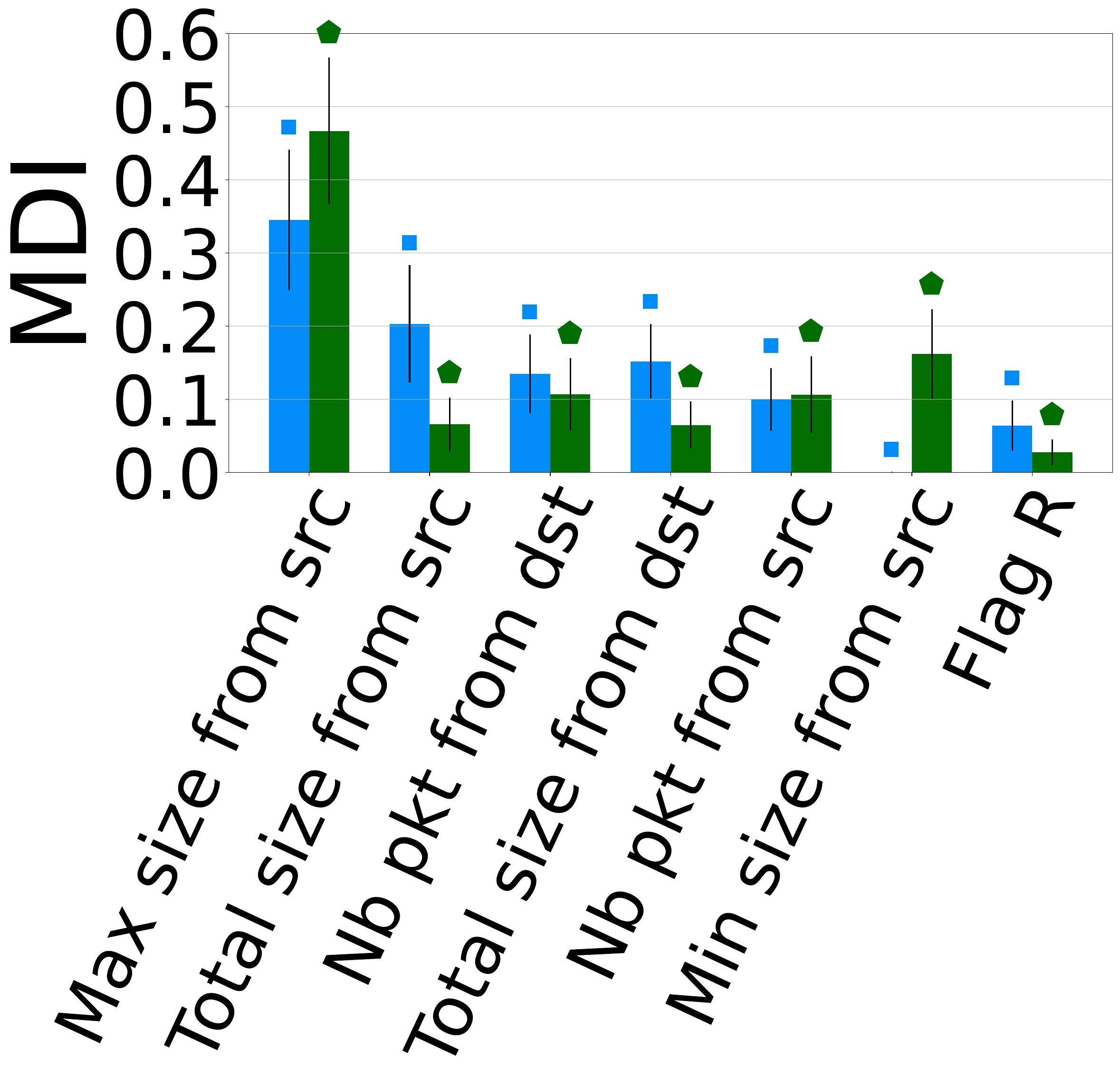}&
			\includegraphics[height=4.5cm]{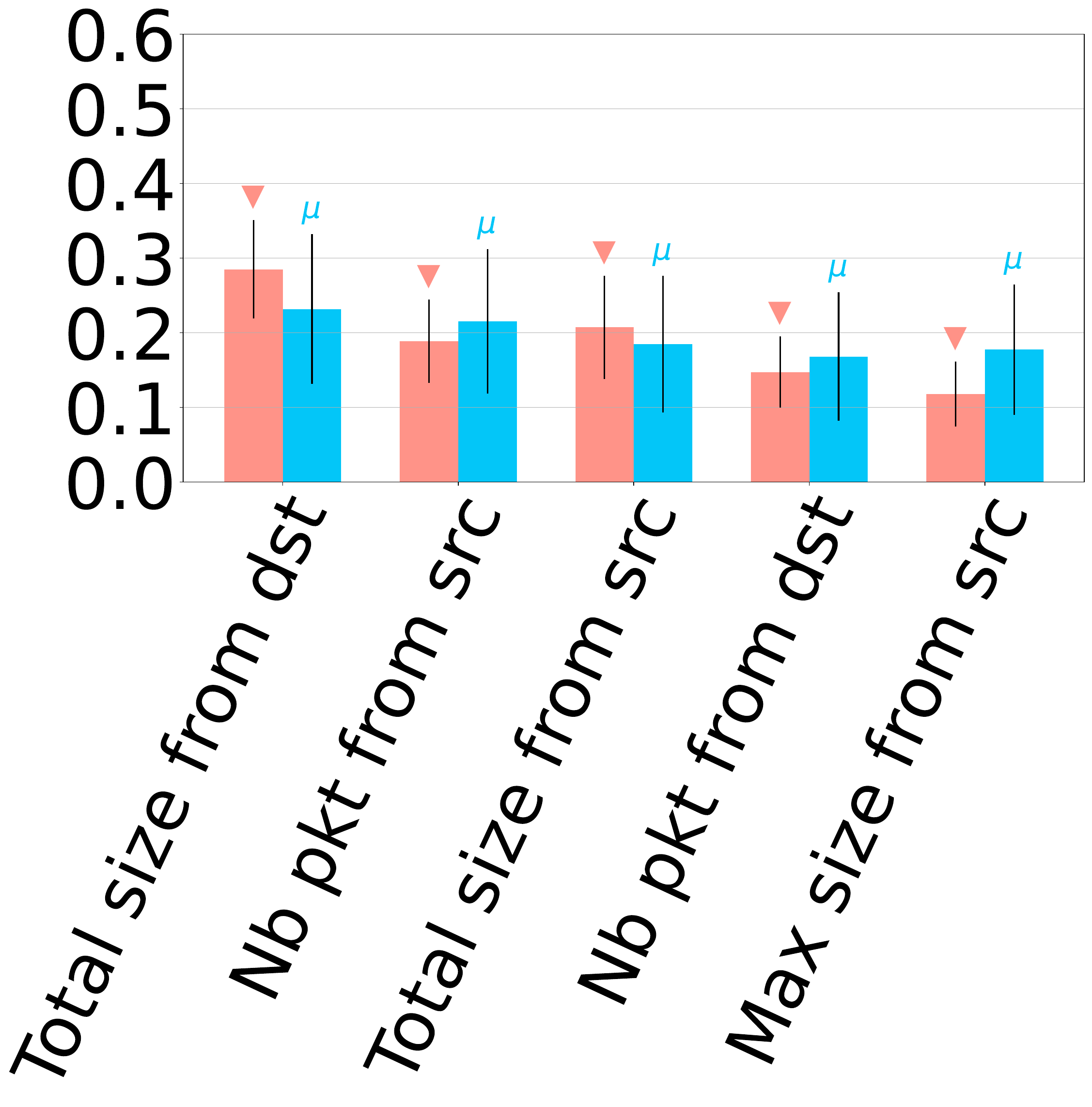}\\
			
		& & \textbf{(d)} & \textbf{(e)}\\
	\end{tabular*}
	\caption{Feature importance based on MDI with a confidence
interval at a 99\% threshold for Netflow v9.
	Only features with a mean value greater than 0.04 are kept.
	Az:~Amazon, D:~Default, Gen:~Generic, jQ:~jQuery, Sb:~Smashburger, pkt:~packet, src:~source (client), dst:~destination (server), R:~Reset.
  }
	\label{fig:fi_noduration}
\end{figure*}

\subsection{Detection of DNS C2 traffic}
\label{subsection:detection_DNS_traffic}
We first evaluate the DNS protocol use case.
The experiments pictured in~\autoref{fig:boxplots} (a) and (b), are approaching the maximum F1 score of 1.
A straightforward observation is the score improvement when Netflow v9
features are used compared to Netflow v5.
This improvement can also be observed in~\autoref{fig:lc_noduration} (d) and (g) as the learning curves reach the maximum $F_1$ score of 1 with ten times less training flows with Netflow v9 features than with Netflow v5 features.
The Mean Decrease in Impurity (MDI) used to measure the importance
for each feature in \autoref{fig:fi_noduration} (a) shows that the total size
of data exchanged is the most important feature whereas the number of packets
is not important at all.
We manually checked that Cobalt Strike is using standard DNS streams composed
mostly of 2 packets which is similar to benign traffic.

\begin{takeaway}{Takeaway}
To summarize our results on DNS C2, the performance of the three methods
are close as they all include the payload size as a feature,
which is visibly the biggest difference between malicious and
benign DNS traffic.
However, our features are easier to collect and more adapted to
production environments.
\end{takeaway}

\subsection{Detection of HTTP(S) C2 traffic linked to an unknown domain}
\label{subsection:detection_unknown_domain}
We then compare both methods when the traffic can not be associated to
a known imitated domain.
Again, we observe a great improvement in the detection capability of
our model using Netflow v9 features compared to Netflow v5.
The detection performance is lower for HTTPS Web browsing-masquerading C2 among generic
traffic than DNS traffic, as pictured in~\autoref{fig:boxplots} (e) to (g).
This is caused by the diversity of traffic in the generic dataset,
having thus more samples similar to the malicious ones.
This lowers the recall and thus the $F_1$ score.

The learning curves corresponding to these experiment
in~\autoref{fig:lc_noduration}, using generic benign traffic,
thus need significantly more training flows to reach an asymptotic value.
It is even hardly reached for HTTPS experiments as shown in~\autoref{fig:lc_noduration} (c), (f) and (i).

The MDI for each feature in experiments with Netflow v5 shows that the total size of data exchanged has the most impact.
We also observe in~\autoref{fig:fi_noduration} (b) and (c) that,
in general with Netflow v9 features, the payload size's extremums from the source have a bigger impact
on the decision than the number of packets.
This is because the size of the payload for a profile has predefined
values during check-ins.
For instance, Amazon profile based Beacon sends data up to 1480 bytes while
the minimum is 20 bytes.
Check-ins without commands exhibit this minimum size, and are thus easy
to identify.
Moreover, the data exfiltration by the Beacon implies that a lot of data is
sent by the source, which contrasts with a benign activity where the client
receives more data than it sends.
Hence, features built on traffic from the source are more crucial
for the final decision.

\begin{takeaway}{Takeaway}
To summarize, the performance of our method with Netflow v9 is similar to
the method proposed by Ramos et al.~\cite{Ramos_ml_cobalt_strike_2022}.
However, the features we use are easier to collect.
Netflow v5 can also be used with lower performance scores.
\end{takeaway}

\subsection{Detection of HTTP(S) C2 traffic linked to a mimicked domain}
\label{subsection:detection_mimicked_domain}
Then we study the case when masquerading traffic can be linked to a known
imitated domain.
Unlike the approach described in Ramos et al.~\cite{Ramos_ml_cobalt_strike_2022},
the method we propose select a specialized model to perform the detection.
The improvement of performance by using Netflow v9 compared to Netflow v5 is
still observed but, more important, the F1 score is greater than with 
Ramos et al.~\cite{Ramos_ml_cobalt_strike_2022} in all experiments as
the median has an equal or higher value and the box plot is tighter, as seen in~\autoref{fig:boxplots}~(h) to (k).

The corresponding learning curves presented in~\autoref{fig:lc_noduration}
show easier learning process for the models to reach the same $F_1$ score 
than models detecting unknown traffic.
They need only several hundreds training flows compared to 
the several thousands flows used in the previous experiments.

For Netflow v5 features, we note that the feature importance difference between
the total size and the number of packets is less significant when the benign
traffic is specific to the domain mimicked by Cobalt Strike.
We suppose that specific benign traffic has a smaller variation in numbers
of packets than generic benign traffic, justifying closer feature
importance values.
Moreover, \autoref{fig:fi_noduration}~(d) shows the impact of the maximum size sent by the client for HTTP experiments.
Again, check-ins without commands stand out when compared to specific benign traffic.

\begin{takeaway}{Takeaway}
	To summarize our results on generated C2 traffic, we obtain good
performances with the Netflow v9 based feature sets.
Furthermore, the use of specific traffic significantly improves
the different scores compared to generic traffic.
Thus, a method that selects a specific model matching the observed traffic
provides the best performances and outperforms the approach proposed by Ramos et al.~\cite{Ramos_ml_cobalt_strike_2022}.
\end{takeaway}

\subsection{Detection of documented real-world Cobalt Strike traffic}
\label{subsection:detection_in_the_wild}
\begin{figure*}[tb]
	\centering
	\setlength\tabcolsep{19.45pt}
	\begin{tabular*}{\textwidth}{ccccc}

		\textbf{11-06} &
		\textbf{01-31} &
		\textbf{05-23} &
		\textbf{07-12} &
		\textbf{10-03}\\

		\includegraphics[height=3.1cm]{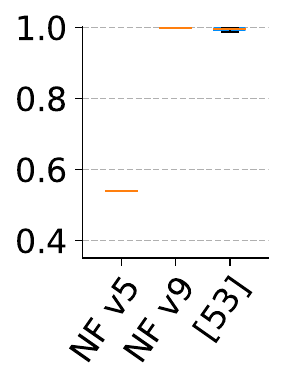} &
		\includegraphics[height=3cm]{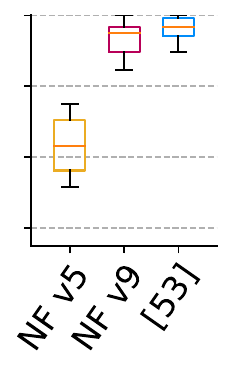} &
		\includegraphics[height=3cm]{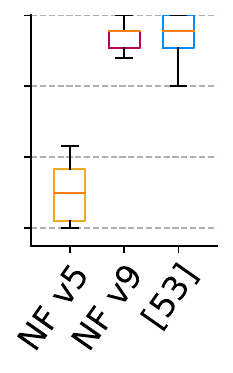} &
		\includegraphics[height=3cm]{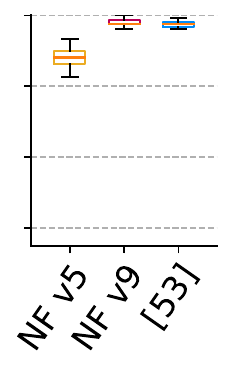} &
		\includegraphics[height=3cm]{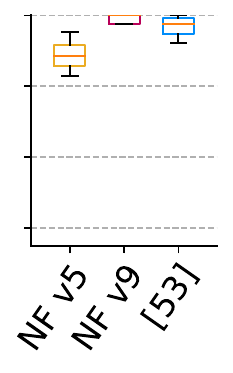}\\
		
		\textbf{(a)} & 
		\textbf{(b)} & 
		\textbf{(c)} & 
		\textbf{(d)} &
		\textbf{(e)} \\
		
	\end{tabular*}
	\caption{$F_1$ score with a confidence interval at a 95\% threshold for detecting masqueraded Cobalt Strike C2 observed in the wild.
	Ramos et al.~\cite{Ramos_ml_cobalt_strike_2022} uses generic HTTP and HTTPS combined benign traffic.
	NF:~NetFlow.
  }
	\label{fig:mta_experiments}
\end{figure*}

Finally, we apply our method to Cobalt Strike C2 traffic collected in 2023~\cite{mta_2023-01-31,mta_2023-05-23,mta_2023-07-12,mta_2023-10-03,mta_2023-11-06}.
To minimize bias, we study traces with at least a hundred Cobalt Strike TCP flows.
This malicious traffic has been collected during real-world
attacks by Malware Traffic Analysis~\cite{mta}, as presented in
\autoref{subsubsection:malicious_traffic}.
The four traces with HTTPS traffic, detailed in \autoref{tab:nb_instances}, contain unencrypted SNIs. However, as they do not correspond to any targeted profile, our method uses several models trained with
generic HTTPS benign traffic, as depicted on the bottom of \autoref{fig:detection}. 

We observe that our method obtains downgraded performances.
This may be occurring because the correct profile is not present inside
training data.
On the contrary, for a trace containing HTTP traffic with a known referer header
corresponding to jQuery~\cite{mta_2023-11-06}, our method uses one model trained
with jQuery HTTP benign traffic.
We then obtain a $F_1$ score of 0.99, as pictured in \autoref{fig:mta_experiments}~(a).
We manually extract and check afterwards that the malleable profile used in 
this trace is, in fact, a jQuery mimicking profile.

To evaluate the capability of Netflow features-based models to detect 
real-world HTTPS Cobalt Strike traffic, we train four models on
the HTTPS selected traces.
We then compare our results to those obtained with the method proposed by
Ramos et al.~\cite{Ramos_ml_cobalt_strike_2022}, as pictured in
\autoref{fig:mta_experiments}~(b) to (e).
Metrics values are presented in appendix in \autoref{tab:metrics},
including precision and recall.
We observe that in most cases, models based on Netflow v9 features
equals or outdo the performances reached by
Ramos et al.~\cite{Ramos_ml_cobalt_strike_2022}.
Thus, Netflow v9 based models are proficient to detect external HTTPS \Cc but Netflow v5 can also be used to create simpler yet effective models.

\begin{takeaway}{Takeaway}
To summarize, our method applied to real-world traces performs well when the profile can be identified (see results for the jQuery profile~\cite{mta_2023-11-06}).
Then, we show that Netflow features-based models' performance roughly equals Ramos et al.~\cite{Ramos_ml_cobalt_strike_2022} and surpass it in some cases.
Thus our method's performance is optimal
when diverse profiles are used for training, as this increases the probability to use a specific model matching the traffic studied.
These experiments show similar results as those in virtualized environment.
\end{takeaway}

\section{Ethics, discussion and future work}
\label{section:discussion}

We acknowledge that inspecting packets has an ethical impact on privacy.
However, we argue that we minimize this impact by searching only for the domain 
name and that our method still provides good results when no domain is
linked to the observed traffic.
Moreover, the traffic generated in our work is not captured from real users' activity.
We identify three main limitations to our work.
First, the number of malicious samples in our experiments is low as we have a limited number of attack scenario to run on our platform.
This prevents us from obtaining optimal results but collecting more information about real use of Cobalt Strike can help to improve our datasets.
Second, the number of malleable profiles studied is still limited while being higher than in most related works.
This is because we limit our experiments to three popular profiles detected in the wild.
This may explain the downgraded performances observed during the evaluation of our method on external HTTPS traces. 
Finally, the generic traffic datasets used are a few years old.
More recent datasets would result to models closer to the current state of
network traffic.
These limitations will be addressed in future work.

We plan to extend this work along three axes.
The first axis is to continue our experiments on other kinds of profiles.
In a \Cc communication, the victim sends more data than it receives.
However, the tested profiles mimic activities where the victim receives more
data than they send, making the detection easier.
By studying the capability to detect data-sending oriented profiles such as
cloud storage uploads or videoconferencing, we suppose it would help to better
estimate the efficiency of other models and improve the coverage offered by
our method.
The second axis is the evaluation of a general performance of our method.
We are also interested in the domain generalization of our method
inside malleable profile groups
(see \autoref{subsubsection:malicious_traffic}), i.e. how well a model
trained on data from a single profile performs on samples from
close-related malleable profiles.
This work will be based on the TLSH-based groups presented in \autoref{tab:profils}.
Finally, the third axis is to extend this work towards other 
frameworks~\cite{sliver_github} and Cobalt Strike extensions that are
used in the wild~\cite{cs_macos}.
We hypothesize that the performance may be preserved as the proposed method is not based on Cobalt Strike's specific parameters.

\section{Conclusion}
\label{section:conclusion}

In this paper, we propose a new network traffic metadata-based machine learning method to detect Cobalt Strike masquerading C2 traffic.
This passive detection method, in opposition with previous work, is based on
unsophisticated but widely used features so it can be easily deployed in
production environments.
It can also adapt the model used to the observed traffic to optimize its performance.
We evaluate the performance of this method on data generated using several widely used Cobalt Strike configurations deployed in a realistic virtualized architecture.
We are the first paper providing such detailed, documented and reproductible evaluation.
We show that a Random Forest model can detect Cobalt Strike masquerading \Cc
with or without encryption.
We also show that this method performs better than the state of art when mimicked websites or services can be identified, and provides similar results to previously proposed approaches when it is not the case.
Finally, we show that our method performance is similar to a previously proposed method on real-world attacks while using more standard features.
The artifacts to reproduce the results of this paper are released at \url{https://github.com/cp-tsp/ares2025-artifacts}.

\section*{Appendix}

\begin{table*}[h]
	\setlength\tabcolsep{4.23pt}
	\caption{Metrics computed and rounded down to one hundredth with a confidence interval at a 95\% threshold of the different experiments in comparison with Ramos et al.~\cite{Ramos_ml_cobalt_strike_2022}.
	Exp: Experiment,
	P: precision,
	R: recall,
	H: HTTP,
	HS: HTTPS,
	D: DNS.}

		\begin{tabular*}{\textwidth}{ldgcdgcdgc}\toprule
	 		\textbf{Exp.} &
			\multicolumn{3}{c}{\textbf{Netflow v5}}&
			\multicolumn{3}{c}{\textbf{Netflow v9}}&
			\multicolumn{3}{c}{\textbf{Ramos et al.~\cite{Ramos_ml_cobalt_strike_2022}}}\\
			
 			\cmidrule(lr){2-4}
	 		\cmidrule(lr){5-7}
			\cmidrule(lr){8-10}
			& $F_1$ & P & R & $F_1$ & P & R & $F_1$ & P & R \\

			Az/Az (HTTP) &
			.98$\pm$.02 & 1$\pm$0 & .96$\pm$.04&
			\textbf{.99$\pm$0} & 1$\pm$0 & .99$\pm$.01&
			& & \\

			Az/Gen (HTTP)&
			.78$\pm$.04 & .85$\pm$.05 & .72$\pm$.06 &
			\textbf{.99$\pm$.01} & .99$\pm$.01 & .99$\pm$.01 &
			\multirow{-2}{*}{.98$\pm$.01} & \multirow{-2}{*}{.99$\pm$.01}	& \multirow{-2}{*}{.98$\pm$.02}\\

			jQ/jQ (HTTP)&
			.99$\pm$0 & .99$\pm$0 & .99$\pm$0 &
			\textbf{1$\pm$0} & 1$\pm$0 & 1$\pm$0 &
			& & \\

			jQ/Gen (HTTP)&
			.98$\pm$0 & .98$\pm$0 & .97$\pm$0 &
			.99$\pm$0 & 1$\pm$0 & .99$\pm$0 &
			\multirow{-2}{*}{.99$\pm$0} & \multirow{-2}{*}{.99$\pm$0} & \multirow{-2}{*}{.99$\pm$0} \\

			Az/Az (HTTPS)&
			.96$\pm$.02 & 1$\pm$0 & .94$\pm$.04 &
			\textbf{.99$\pm$.01} & .99$\pm$.01 & .98$\pm$.01 &
			& & \\

			Az/Gen (HTTPS)&
			.84$\pm$.05 & .88$\pm$.05 & .81$\pm$.06 &
			.92$\pm$.02 & .99$\pm$.01 & .87$\pm$.05 &
			\multirow{-2}{*}{.95$\pm$.02} & \multirow{-2}{*}{1$\pm$0} & \multirow{-2}{*}{.91$\pm$.04}\\
		
			Sb/Sb (HTTPS)&
			.99$\pm$.01 & .98$\pm$.02 & 1$\pm$0 &
			\textbf{1$\pm$0} & 1$\pm$0 & 1$\pm$0 &
			& & \\

			Sb/Gen (HTTPS)&
			.84$\pm$.04 & .88$\pm$.05 & .81$\pm$.07 &
			.95$\pm$.03 & 1$\pm$0 & .91$\pm$.06 &
			\multirow{-2}{*}{.95$\pm$.03} & \multirow{-2}{*}{1$\pm$0} &
			\multirow{-2}{*}{.91$\pm$.06}\\
		
			D/Gen (HTTPS)&
			.80$\pm$.05 & .87$\pm$.05 & .75$\pm$.07 &
			\textbf{.97$\pm$.01} & 1$\pm$0 & .95$\pm$.02 &
			.96$\pm$.02 & 1$\pm$0 & .94$\pm$.04 \\

			D/Gen (D)&
			.96$\pm$0 & .95$\pm$0 & .97$\pm$0 &
			\textbf{.99$\pm$0} & .99$\pm$0 & .99$\pm$0 &
			\textbf{.99$\pm$0} & .99$\pm$0 & .99$\pm$0\\
		
			Az/Gen (D)&
			.96$\pm$0 & .96$\pm$0 & .97$\pm$0	&
			\textbf{.99$\pm$0} & .99$\pm$0 & .99$\pm$0 &
			\textbf{.99$\pm$0} & .99$\pm$0 & .99$\pm$0\\

			\midrule

			01-31 (HTTPS)&
			.60$\pm$.10 & .68$\pm$.11 & .63$\pm$.09 &
			.92$\pm$.05 & 1$\pm$0 & .88$\pm$.06 &
			\textbf{.94$\pm$.05}& 1$\pm$0 & .89$\pm$.05 \\

			05-23 (HTTPS)&
			.49$\pm$.05 & .54$\pm$.08 & .48$\pm$.09 &
			.93$\pm$.04 & .99$\pm$.01 & .87$\pm$.07 &
			\textbf{.94$\pm$.04} & 1$\pm$0 & .89$\pm$.05 \\

			07-12 (HTTPS)&
			.88$\pm$.02 & .87$\pm$.03 & .88$\pm$.02 &
			\textbf{.98$\pm$0} & 1$\pm$0 & .96$\pm$.01 &
			.97$\pm$0 & 1$\pm$0 & .95$\pm$.02 \\

			10-03 (HTTPS)&
			.88$\pm$.02 & .91$\pm$.03 & .87$\pm$.04 &
			\textbf{.98$\pm$0} & 1$\pm$0 & .97$\pm$.02 &
			.96$\pm$.02 & 1$\pm$0 & .93$\pm$.04 \\

			11-06 (HTTP)&
			.56$\pm$.04 & .99$\pm$0 & .40$\pm$.06 &
			\textbf{.99$\pm$0} & 1$\pm$0 & .99$\pm$0 &
			\textbf{.99$\pm$0} & 1$\pm$0 & .99$\pm$0 \\
			\toprule

 		\textbf{Exp.} &
		\multicolumn{3}{c}{\textbf{Netflow v5 extended}}&
 		\multicolumn{3}{c}{\textbf{Netflow v9 extended}}&
 		\multicolumn{3}{c}{\textbf{Ramos et al.~\cite{Ramos_ml_cobalt_strike_2022}}}
		\\
 		\cmidrule(lr){2-4}
 		\cmidrule(lr){5-7}
		\cmidrule(lr){8-10}

 		& $F_1$ & P & R &
 		$F_1$ & P & R &
		$F_1$ & P & R\\

		Az/Az (HTTP) & 
		.98$\pm$.01 & 1$\pm$0 & .97$\pm$.03 &
		\textbf{1$\pm$0} & 1$\pm$0 & 1$\pm$0 &
    		& & \\
		
		Az/Gen (HTTP)&
	 	.82$\pm$.03	& .88$\pm$.06 & .78$\pm$.05 &
		\textbf{.98$\pm$.01} & .98$\pm$.01	& .97$\pm$.02 &
		\multirow{-2}{*}{.98$\pm$.01} & \multirow{-2}{*}{.99$\pm$.01}	& \multirow{-2}{*}{.98$\pm$.02}\\
		
		jQ/jQ (HTTP)&
		\textbf{.99$\pm$0} & .99$\pm$0 & .99$\pm$0 &
		\textbf{.99$\pm$0} & 1$\pm$0 & .99$\pm$0 &
		 & & \\
		
		jQ/Gen (HTTP)&
		.98$\pm$0 & .98$\pm$0 &	.97$\pm$0 & 
		\textbf{.99$\pm$0} & .99$\pm$0 & .99$\pm$0 &
		\multirow{-2}{*}{\textbf{.99$\pm$0}} & \multirow{-2}{*}{.99$\pm$0} & \multirow{-2}{*}{.99$\pm$0} \\
			
		Az/Az (HTTPS)&
	 	.97$\pm$.01	& .97$\pm$.03	& .97$\pm$.02 &
		\textbf{.99$\pm$0} & .99$\pm$.01 & 1$\pm$0 &
		 & & \\
		
		Az/Gen (HTTPS)&
 		.86$\pm$.03 &	.91$\pm$.02	& .81$\pm$.07 &
		.91$\pm$.03 & .98$\pm$.01	& .85$\pm$.06 &
		\multirow{-2}{*}{.95$\pm$.02} & \multirow{-2}{*}{1$\pm$0} & \multirow{-2}{*}{.91$\pm$.04} \\
		
		Sb/Sb (HTTPS)&
		 .99$\pm$.01 & .98$\pm$.02 & 1$\pm$0 &
		 \textbf{1$\pm$0} & 1$\pm$0 & 1$\pm$0 &
		 & & \\
		
		Sb/Gen (HTTPS)&
		.86$\pm$.05	& .92$\pm$.05 & .82$\pm$.08 &
		.94$\pm$.02	& 1$\pm$0 & .89$\pm$.05 &
		\multirow{-2}{*}{.95$\pm$.03} & \multirow{-2}{*}{1$\pm$0} & \multirow{-2}{*}{.91$\pm$.06}\\
		
		D/Gen (HTTPS)&
 		.80$\pm$.05	& .88$\pm$.05 & .74$\pm$.07 &
		\textbf{.96$\pm$.01} & 1$\pm$0 & .93$\pm$.02 &
		.96$\pm$.02 & 1$\pm$0 & .94$\pm$.04 \\
		
		D/Gen (D)&
		.96$\pm$0 & .95$\pm$0 & .97$\pm$0 &
		\textbf{.99$\pm$0} & .99$\pm$0 & .99$\pm$0 &
		\textbf{.99$\pm$0} & .99$\pm$0 & .99$\pm$0\\
		
		Az/Gen (D)&
		.96$\pm$0 & .96$\pm$0 & .97$\pm$0 &
		\textbf{.99$\pm$0} & .99$\pm$0 & .99$\pm$0 &
		\textbf{.99$\pm$0} & .99$\pm$0 & .99$\pm$0 \\

		\bottomrule
 	\end{tabular*}
	\label{tab:metrics}
\end{table*}
\newpage
\bibliographystyle{splncs04}
\bibliography{biblio}

\end{document}